\documentclass[sn-nature]{sn-jnl}

\usepackage{graphicx}%
\usepackage{multirow}%
\usepackage{amsmath,amssymb,amsfonts}%
\usepackage{amsthm}%
\usepackage{mathrsfs}%
\usepackage[title]{appendix}%
\usepackage{xcolor}%
\usepackage{textcomp}%
\usepackage{manyfoot}%
\usepackage{booktabs}%
\usepackage{algorithm}%
\usepackage{algorithmicx}%
\usepackage{algpseudocode}%
\usepackage{listings}%
\usepackage{ragged2e}
\usepackage{gensymb} 
\usepackage{xspace} 

\newcommand\vphi[1]{$\upsilon_{\phi}$}
\newcommand\dvphi[1]{$\delta\upsilon_{\phi}$}
\newcommand\vz[1]{$\upsilon_{\rm z}$}

\newcommand{\twCOfull}{$^{12}$CO\,$J=2-1$\,}
\newcommand{\twCO}{$^{12}$CO\,}

\newcommand{\farcs}{{.}''}

\newcommand{\wisa}{WISPIT~2\xspace}
\newcommand{\wisb}{WISPIT~2b\xspace}
\newcommand{\wisc}{WISPIT~2c\xspace}

\begin{document}


\title[Planet gaps in WISPIT 2]{Planetary gas gaps and kinematic signatures in the planet forming disk around WISPIT-2}

\author[1]{{\fnm{Nienke} \sur{van der Marel} }\email{nmarel@strw.leidenuniv.nl}}
\equalcont{These authors contributed equally to this work.}

\author[2]{{\fnm{Jochen} \sur{Stadler} }\email{jochen.stadler@eso.org}}
\equalcont{These authors contributed equally to this work.}

\author[3]{{\fnm{Andres F.} \sur{Izquierdo} }\email{aizquierdocartag@ufl.edu}}

\author[4]{{\fnm{Ruobing} \sur{Dong} }\email{rbdong@pku.edu.cn}}

\author[5]{{\fnm{Bertram} \sur{Bitsch} }\email{bbitsch@ucc.ie}}

\author[6]{{\fnm{Jake} \sur{Byrne} }\email{byrnej77@tcd.ie}}

\author[1]{{\fnm{Richelle} \sur{van Capelleveen} }\email{capelleveen@strw.leidenuniv.nl}}

\author[7,8]{{\fnm{Chloe} \sur{Lawlor} }\email{c.lawlor13@universityofgalway.ie}}

\author[7,8]{{\fnm{Christian} \sur{Ginski} }\email{christian.ginski@universityofgalway.ie}}

\author[1]{{\fnm{Matthew} \sur{Kenworthy} }\email{kenworthy@strw.leidenuniv.nl}}

\affil[1]{\orgdiv{Leiden Observatory}, \orgname{Leiden University}, \orgaddress{\street{P.O. Box 9513}, \city{Leiden}, \postcode{2300 RA}, \country{The Netherlands}}}

\affil[2]{\orgdiv{European Southern Observatory}, \orgname{ESO}, \orgaddress{\street{Karl-Schwarzschild-Str. 2}, \city{Garching bei München}, \postcode{85748}, \country{Germany}}}

\affil[3]{\orgdiv{Department of Astronomy}, \orgname{University of Florida}, \orgaddress{\street{211 Bryant Space Science Center}, \city{Gainesville}, \postcode{FL 32611}, \country{USA}}}

\affil[4]{\orgdiv{Kavli Institute for Astronomy and Astrophysics}, \orgname{Peking University}, \orgaddress{\street{5 Yiheyuan Road}, \city{Beijing}, \postcode{100871}, \country{People's Republic of China}}}

\affil[5]{\orgdiv{Department of Physics}, \orgname{University College Cork}, \orgaddress{\street{College Road}, \city{Cork}, \postcode{T12K8AF}, \country{Ireland}}}

\affil[6]{\orgdiv{School of Physics}, \orgname{Trinity College Dublin, The University of Dublin}, \orgaddress{\street{College Green}, \city{Dublin 2}, \postcode{D02 PN40}, \country{Ireland}}}

\affil[7]{\orgdiv{Physics, School of Natural Sciences}, \orgname{University of Galway}, \orgaddress{\street{University Road}, \city{Galway}, \postcode{H91 TK33}, \country{Ireland}}}

\affil[8]{\orgdiv{Ryan Institute}, \orgname{University of Galway}, \orgaddress{University Road}, \city{Galway}, \postcode{H91 TK33}, \country{Ireland}}



\abstract{
The giant planet formation process in disks of gas and dust surrounding young stars is observationally still poorly constrained.  
Direct detection of protoplanets within the disk remains limited with current facilities. Indirect observational evidence of protoplanets through substructures or deviations from Keplerian rotation in the gas remains ambiguous in absence of detected planets.
The lack of the combined detection of substructures and their corresponding planets makes it challenging to connect the planet formation process to the host environment.
Observations with the Atacama Large Millimeter Array (ALMA) reveal clear detections of gas gaps and kinematic signatures in the \wisa protoplanetary disk that are cospatial with the locations of previously detected giant protoplanets.
The newly identified gas gaps, observed in both the $^{12}$CO integrated intensity map and its rotation curve, correspond with the previously identified gaps in scattered light. Their morphologies are shown to be consistent with eccentric gaps, as expected from planet-disk interaction models of massive planets. The lack of eccentricity in the outer disk can be explained if a third planet is present in the system. 
The observations of \wisa provide a long-sought empirical bridge between kinematic signatures, gas gaps and giant planet formation in disks.
}


\maketitle

\section{Main Article}\label{sec1}


%
Exoplanets are now known to be ubiquitous throughout the Galaxy \citep{ZhuDong2021}, but understanding the planet formation mechanism in disks of gas and dust surrounding young stars remains a long-standing problem in astronomy \citep{Raymond2022}. Observing protoplanets directly in their host environment while they are forming is the ideal way to understand their formation environment and timescale. Despite numerous attempts, the direct detection of protoplanets in disks remains limited to a handful of cases \citep[e.g.][]{Keppler2018,Haffert2019,Currie2022}. 
In the past decade, many protoplanetary disks have shown evidence for gaps in their dust and gas distributions \citep[e.g.][]{vanderMarel2016-isot,Andrews2020}, and some exhibit kinematic deviations from Keplerian rotation in molecular line observations \citep[e.g.][]{Teague2018, Izquierdo_ea_2022, Pinte_ea_PPVII}. These features are often attributed to forming planets, but remain indirect and ambiguous because they can also arise from other disk processes, including hydrodynamic and gravitational instabilities, winds, shadowing effects, disk warps, or density gaps unrelated to planets  \citep[e.g.][]{Bae2023}.
Connecting disk substructures to the planet formation process thus remains highly challenging, in the absence of observations where both protoplanets and substructures have been directly identified.

\begin{figure*}[!ht]
    \centering
    \includegraphics[width=\textwidth]{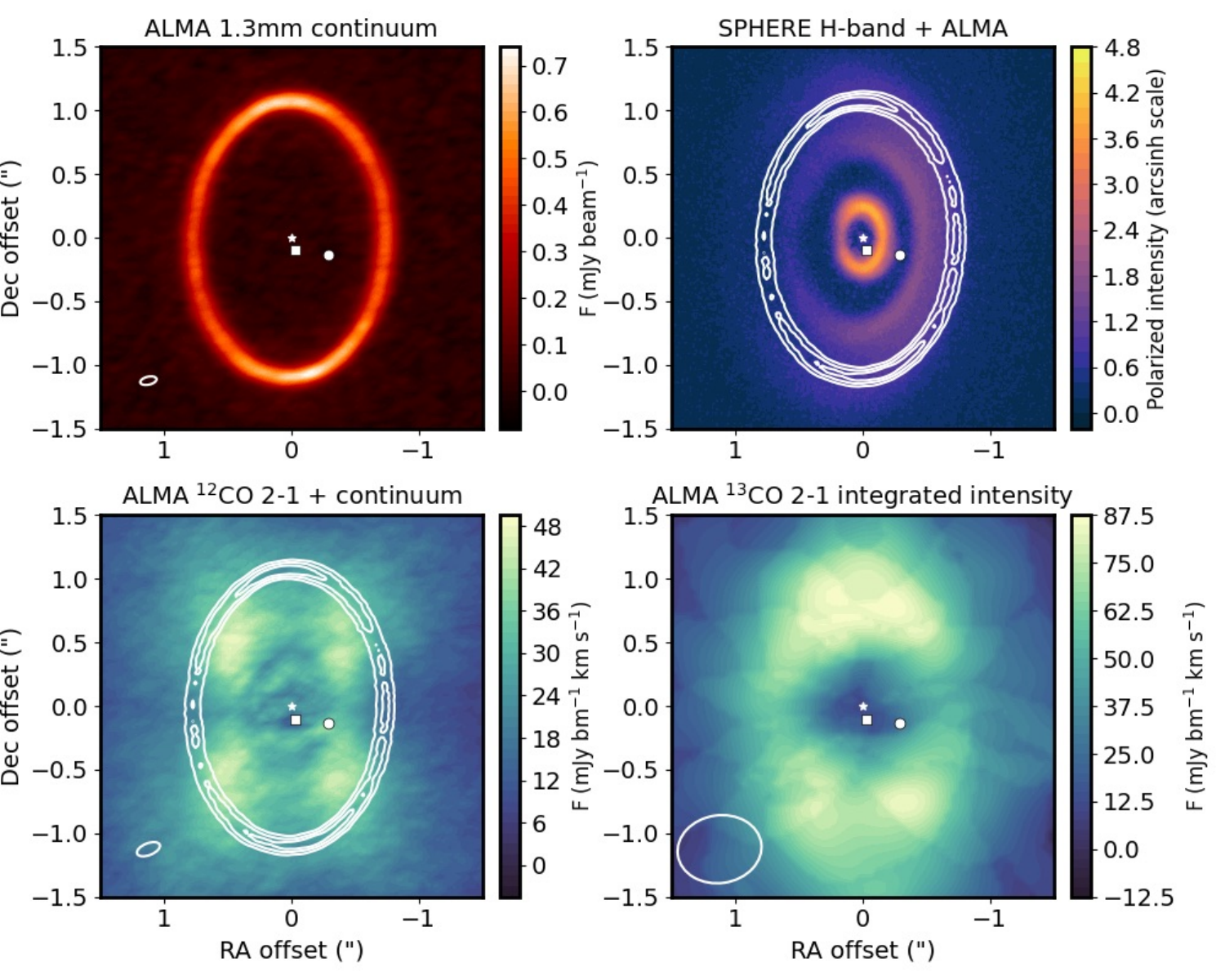}
    \caption{Overview of the concatenated ALMA images of the 1.3~mm continuum, $^{12}$CO 2--1, and $^{13}$CO 2--1 integrated intensity maps, compared with the $H$-band SPHERE image of \wisa in dual-beam polarimetric imaging (DPI) mode. The star is indicated with a white star sign and the protoplanets with a white square and circle. The ALMA continuum contours in the upper right and lower left panels are shown at 20,30,40 times the rms. The beam of each ALMA image is indicated in the lower left corner. 
    }
    \label{fig:mainimages}
\end{figure*}

%
Using new observations from the Atacama Large Millimeter/submillimeter Array (ALMA), we report the detection of gas gaps as well as deviations from Keplerian motion in the protoplanetary disk around \wisa \citep{vanCapelleveen2025,Close2025,Lawlor2026}, including a localized perturbation in the gas velocity field. These signatures are found at the exact locations where (accreting) protoplanets have been previously identified (Figure \ref{fig:mainimages}). We demonstrate that the gas and dust observations are consistent with the predictions from planet-disk interaction models, in which forming giant planets sculpt their surrounding disk (see Methods). The \wisa system, located at a distance of 133 pc, consists of a young Solar-type star ($T_{\rm eff}$=4400 K) surrounded by a dusty disk with multiple dust gaps in scattered light as imaged with VLT/SPHERE, where two massive protoplanets have been directly detected in two of these gaps \citep{vanCapelleveen2025, Close2025, Lawlor2026}. The first discovered planet \wisb was confirmed through multi-epoch, multi-wavelength SPHERE observations at 1.6 and 2.2 $\mu$m consistent with a 5 $M_{\rm Jup}$ planet at 57$^{+8}_{-3}$ au distance from the star, as well as localized H$\alpha$ emission, indicating that the planet is actively accreting gas from its vicinity. \wisc was initially discovered at 3.5 $\mu$m and confirmed with VLTI/GRAVITY and VLT/SPHERE observations, revealing a $\sim$9 $M_{\rm Jup}$ planet at $\sim$14 au. The star and its planets have been estimated to be 5 Myr of age.

%
%
While the inferred planet locations coincide with the gaps identified in scattered light emission \citep{vanCapelleveen2025}, the outer planet \wisb alone cannot explain the location of the millimeter-dust ring detected in previous ALMA continuum observations \citep{Facchini2026}. While multiple planets can open a wider and shallower common gap, the outer gap edge remains largely controlled by the outermost planet \citep{DuffellDong2015}; thus, adding the inner planet \wisc is unlikely to resolve the large separation between \wisb and the millimeter ring. Generally, a millimeter-dust ring is expected to arise from dust trapping at the outer edge of a planet gap in a so-called pressure bump \citep{Paardekooper2004, Pinilla2012b}, but here the ring is located at more than 2 times the outer planet orbital radius. 
However, a planet’s gravitational perturbation produces distinct observable signatures in the gas disc: (1) it modifies the gas surface density by repelling gas from its orbit and opening a gap, and (2) it perturbs the gas velocity field, producing deviation from Keplerian rotation \citep{Paardekooper2023}. 
Scattered light is only indirectly sensitive to the gas in the disk by tracing the small dust grains at the disk surface that are well coupled to the gas and are subject to illumination and shadowing effects \citep{Benisty2023}.

\begin{figure*}[!ht]
    \centering
    \includegraphics[width=\textwidth]{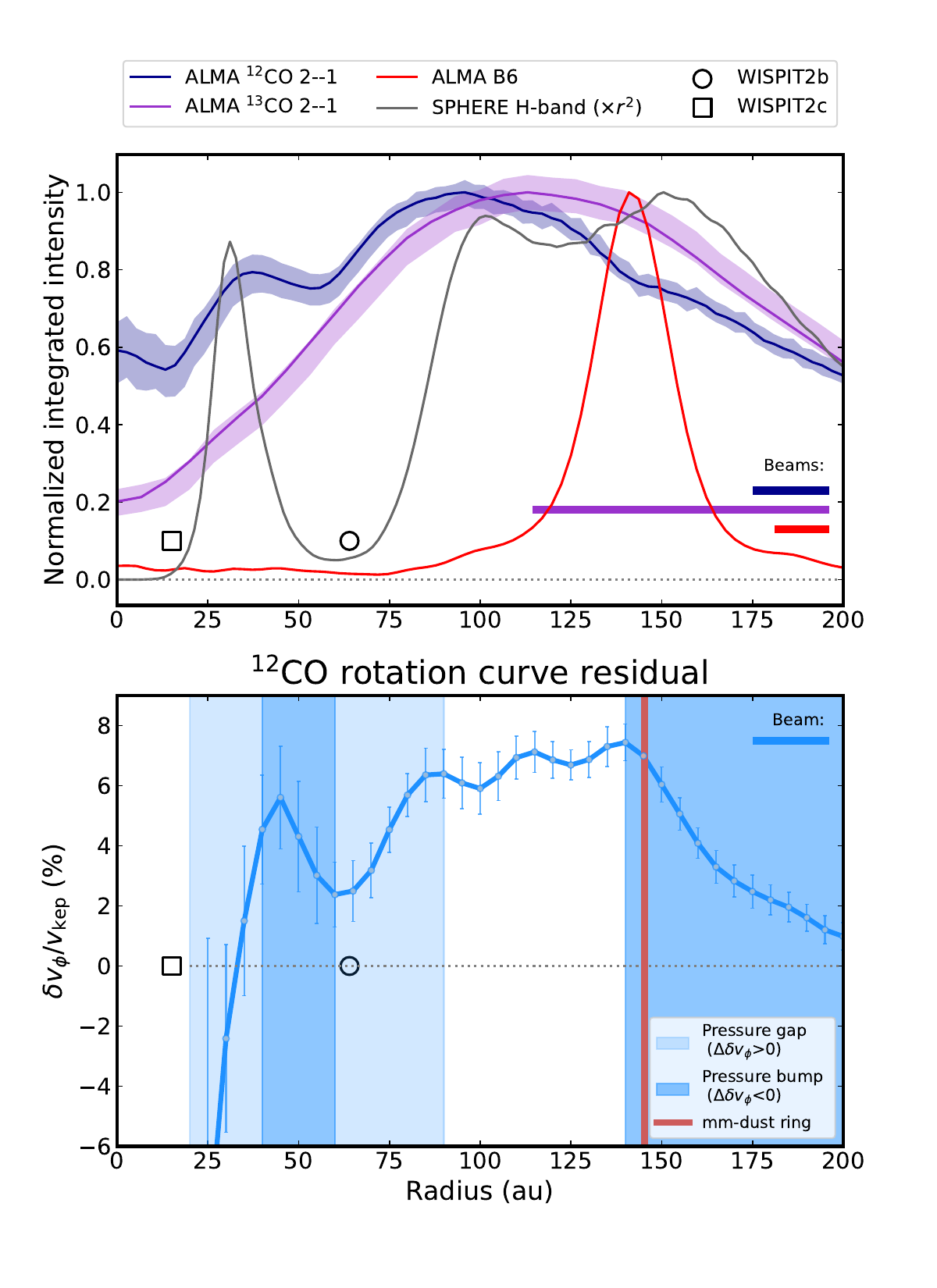}

    \caption{Radial profiles of CO and dust continuum emission. (Top) Normalized intensity radial profiles of the ALMA 1.3\,mm continuum, $^{12}$CO 2--1, and $^{13}$CO 2--1 integrated intensity maps, compared with the DPI $H$-band SPHERE image of \wisa (all from Figure \ref{fig:mainimages}). The $^{12}$CO radial profile has been extracted using the derived emitting surface profile from the Keplerian disk model (see Methods). The colored regions indicate the standard deviations within the radial bins. The average beam size for each curve is indicated on the right to help gauge the size of the observed spatial structures. (Bottom) Deviations from Keplerian rotation in the $^{12}$CO 2--1 data from the $\upsilon_{\phi}$ curve after subtraction of the Keplerian model profile, divided by the Keplerian velocity. The peak of the millimeter-dust ring is indicated with a red vertical line. In both panels, the protoplanet positions are indicated with a square and circle, as in Figure \ref{fig:mainimages}. Colored regions indicate whether the curve increases with radius (consistent with a pressure gap) or decreases with radius (consistent with a pressure bump). The protoplanets overlap with pressure gap regions, while the millimeter-dust ring overlaps with a pressure bump region. Variations between 90 and 140 au are tentative, and thus no gradients are indicated.}
    \label{fig:radialprofiles}
\end{figure*}

%
%
The ALMA $^{12}$CO 2--1 intensity map (Figure \ref{fig:mainimages}) and its azimuthally averaged radial profile (Figure \ref{fig:radialprofiles}), observed at $\sim0\farcs$15 or 20 au resolution, shows clear evidence for drops in the emission around $\sim$15 and $\sim$60 au radius, corresponding to the planet locations. The radial profile has been extracted using the derived emitting surface profile from the Keplerian disk model (see Methods). The gas disk itself extends out to 600 au. The $^{13}$CO 2--1 emission (taken at a lower spatial resolution of 0$\farcs$6 or 80 au) shows a single gas cavity with a radius of 100 au, indicating that the line emission inside this radius is partially optically thin. C$^{18}$O 2--1 emission was not detected, implying that the disk's gas mass may be relatively low \citep{Miotello2014}. 
The 1.3\,mm continuum emission (beam size 0$\farcs$13$\times$0$\farcs$06) reveals a narrow circular dust ring, with a peak at 144 au. Intensity modeling of the continuum visibility data (see Methods) reveals an inner pedestal of millimeter emission at 80 au, which was not seen in previous imaging of the millimeter disk, likely due to lack of short baseline data \citep{Facchini2026}. 

\begin{figure*}[!ht]
    \centering
    \includegraphics[width=0.9\textwidth]{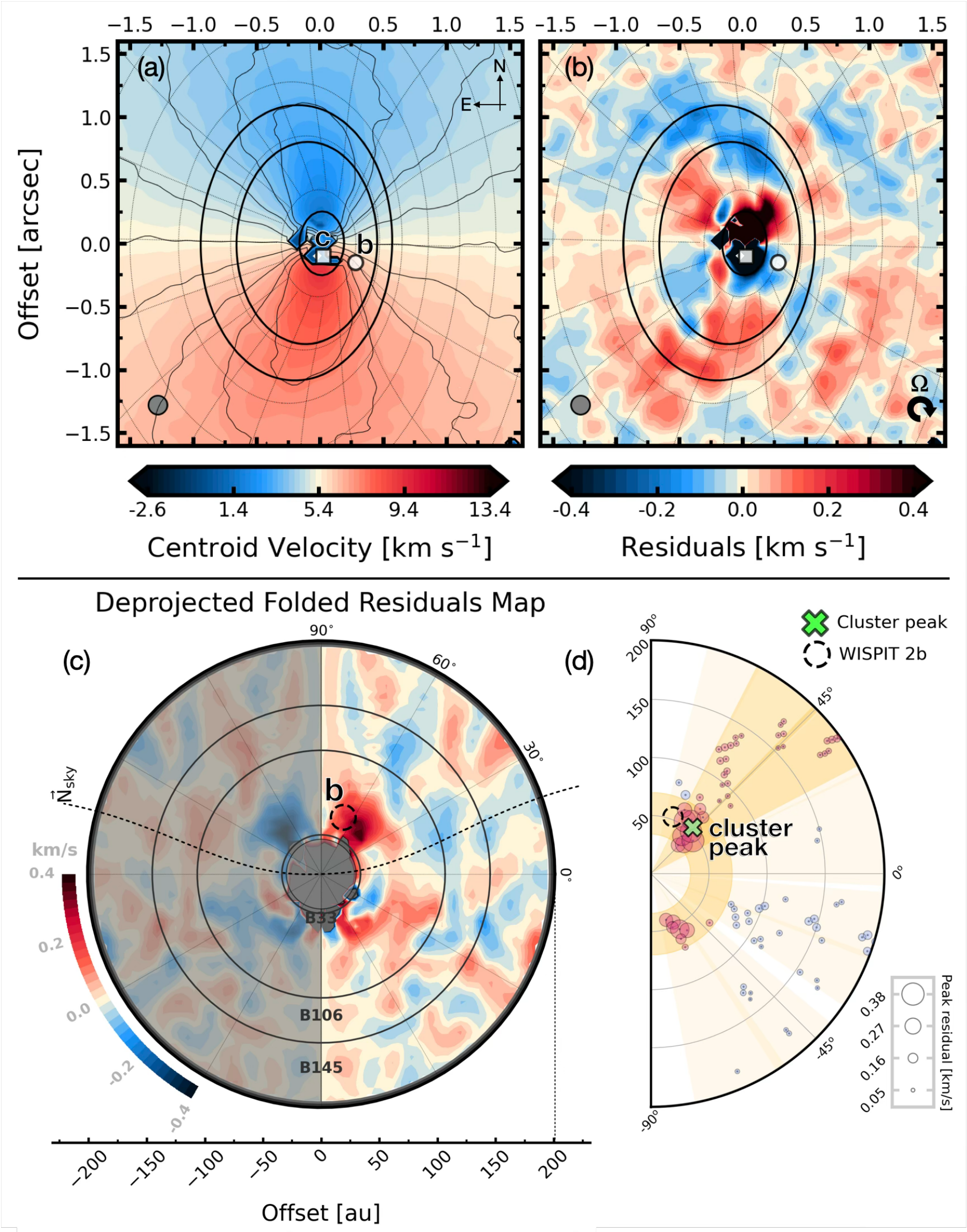}
    \caption{Kinematic detection of \wisb in ALMA $^{12}$CO 2--1 data. a) Gaussian line-of-sight centroid velocity map. b) Residuals after subtracting the Keplerian model from the velocity moment map. In the upper panels, the beam-size is shown as gray circles in the lower left corners, and the locations of planets \wisa b and c are highlighted by white circles and squares, respectively. The black ellipses indicate the locations of IR rings (B33 and B106) and the peak of the continuum (B145), all projected onto the \twCO{} emission surface. c) Resulting residual map after additive folding the velocity residual map along the disk minor axis to remove the axisymmetric rotational components. The plot is in the disk reference frame; North is indicated by the dashed black line. d) Map showing the peak velocity residual points detected in the folded map, scaled in size by the magnitude of their deviation. The detected significant clusters in azimuth and radius are highlighted in dark yellow wedge and annulus, respectively (see Methods for details). The inferred localized cluster peak is shown as a green lime cross. Its cluster center is located within one beam size of the black circle that indicates the current on-sky position of \wisb deprojected to the disk midplane.}
    \label{fig:kinematic_detection}
\end{figure*}

%
%
A Keplerian disk model is fit to the $^{12}$CO channel maps using state-of-the-art modeling tools (see Methods), and is found to be consistent with a rotating disk inclined at 46$^{\circ}$, orbiting a central mass of 1.37$\pm0.02$ $M_{\odot}$. This estimate is 20\% higher than the previously derived stellar mass of \wisa of 1.08$^{+0.06}_{-0.17}$ M$_{\odot}$ based on stellar isochrones \citep{vanCapelleveen2025}, indicating either a different age, larger uncertainties on the isochrones, or a close-in binary companion. After subtracting the Keplerian model and removing the axisymmetric rotational component associated with the disk radial pressure gradient, we identify localized velocity perturbations in the resulting folded residual map (see Figure~\ref{fig:kinematic_detection}). By detecting peak residual points and performing a clustering analysis as in \citep[][detailed in Methods]{Izquierdo2021}, we find statistically significant, spatially localized deviations from Keplerian motion close to (within one beam size) the location of \wisb. These signatures are consistent with a combination of velocity perturbations tracing the wake of spiral density waves, meridional flows and/or circumplanetary disk motion, which are all expected in the vicinity of a (forming) planet \citep{PerezS2015,Izquierdo2021,Pinte_ea_PPVII}. Localized deviations could not be detected at the location of \wisc at the observed spatial resolution.

\begin{figure*}[!ht]
    \centering
    \includegraphics[width=\textwidth,trim=50 50 50 50]{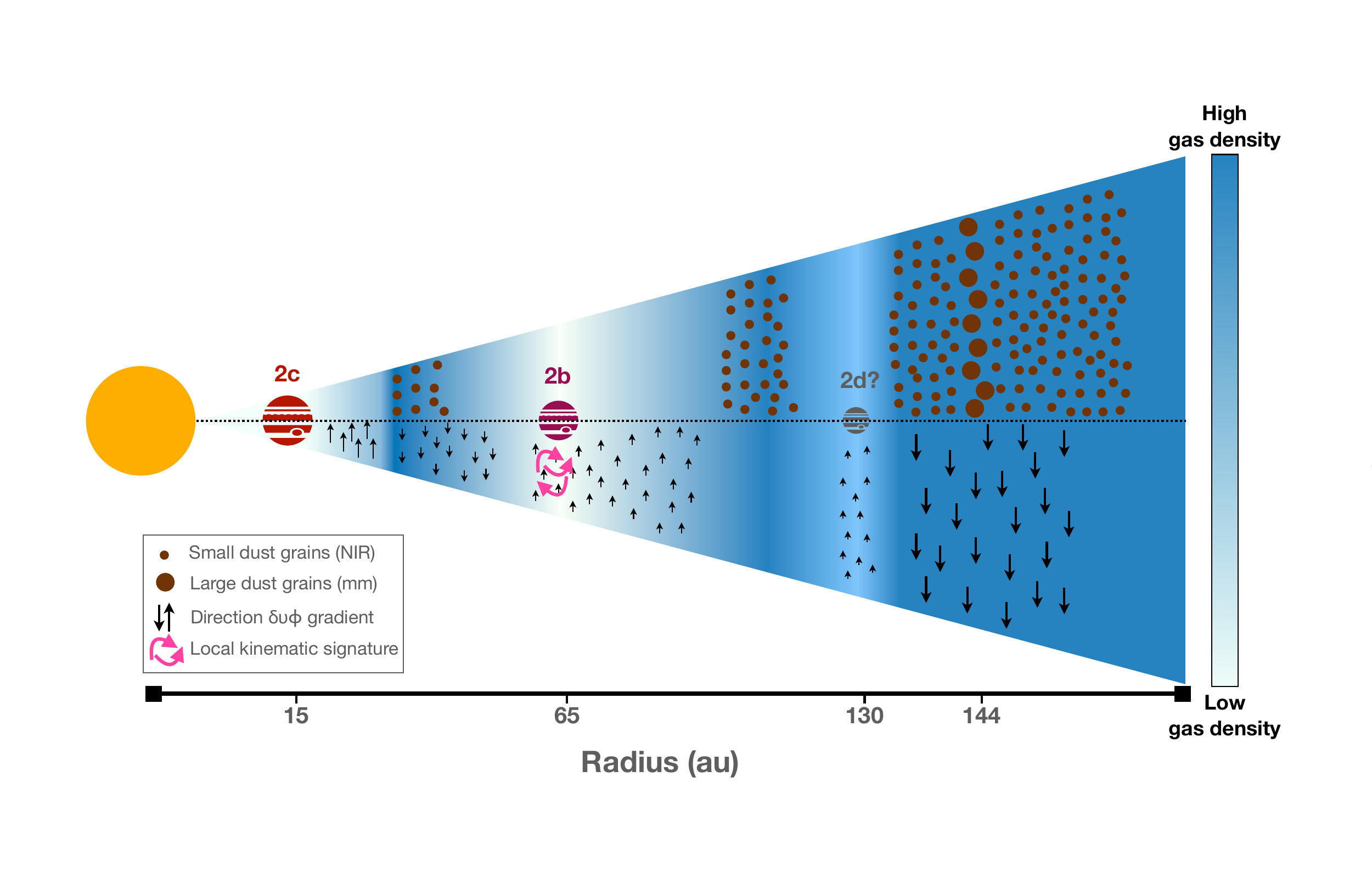}
    \caption{Graphic of the derived gas and dust distribution in the \wisa disk. Gas surface density is shown in blue, where lighter regions indicate drops in density. The top half shows the dust grains in brown circles (small: scattered light, large: millimeter continuum).
    The bottom half shows arrows indicating the direction of the $\delta \upsilon _{\phi}$ pressure gradient, where upward arrows are consistent with a pressure minimum, and downward arrows are consistent with a pressure bump. 
    Magenta curved arrows indicate the local kinematic signature cospatial with \wisb (Figure \ref{fig:kinematic_detection}).  
    }
    \label{fig:graphic}
\end{figure*}

%
%
By disentangling the three line-of-sight velocity components of the $^{12}$CO emission ($v_z$, $v_r$ and $v_{\phi}$, see Methods for details), we have also identified deviations from Keplerian rotation in the $v_{\phi}$ curve (Figure \ref{fig:radialprofiles}), which are signaling the presence of pressure troughs (increasing \dvphi{} with radius) or pressure bumps (decreasing \dvphi{} with radius) \citep{Teague2018, Rosotti_ea_2020}. Positive gradients are found at radii $<$40 au and 55-90 au, coinciding with planet 2c and 2b, whereas a negative gradient is seen at 135-170 au, coinciding with the millimeter dust ring. Tentative evidence for a negative gradient at 40-55 au hints at a second pressure bump between planet 2c and 2b, but no millimeter-dust ring is detected here. This indicates either that the intermediate pressure bump is too shallow to trap any dust, or that the outer pressure bump at 144 au has efficiently trapped all millimeter dust \citep{Pinilla2012b}. Figure \ref{fig:graphic} shows the derived distribution of gas and dust in \wisa, including the direction of the pressure gradient.

%
%
From both the $^{12}$CO intensity profile and the \dvphi{} gradients it is evident that the gaps are radially wider than typical gap widths as predicted by planet-disk interaction models \citep{Crida2006}, as well as radially non-axisymmetric (Figure~\ref{fig:radialprofiles}): the outer gap edge lies about twice as far from the planet as the inner gap edge. Such a gap morphology is qualitatively consistent with eccentric gaps. 
Meanwhile, we caution that projection or optical depth effects may contribute to the apparent radial asymmetry. Nevertheless, the eccentric gap interpretation is supported by independent evidence from the dust morphology and planet orbits: re-analysis of the scattered light image using the orientation of the outer disk as derived from the Keplerian disk model confirms that the $\mu$m-dust rings at 34 and 107 au just outside of these gaps are consistent with eccentricity of $\sim$0.13 (see Methods). A re-fit of the planet orbital data finds mean orbital separations of 15$\pm$2 and 65$\pm4$ au, respectively (see Methods), and indicates that a non-zero eccentricity is likely for both planets. Unfortunately, the spatial resolution of the CO data is insufficient to confirm eccentricity of the gas gaps directly.

Planets are expected to induce eccentricity in their surrounding disk via Lindblad resonances when sufficiently massive \citep{Kley2006,Bitsch2013}. This can induce eccentricity in the planet's orbit through gas-disk interactions \citep{dAngelo2006}, leading to a wide, asymmetric gas gap that is wider exterior to the planet's orbit but shallower in depth \citep{Teyssannier2017,Muley2019,Padgett2026}. Both protoplanets in the \wisa system exceed the mass-ratio threshold at which eccentricity is expected to be induced \citep{Kley2006}. However, the gas kinematics in the outer disk show no signatures of disk eccentricity, as those predicted by hydrodynamic simulations \citep{Kuo2022, Ragusa2024}. Additionally, the ALMA millimeter-dust ring at 144 au is consistent with a circular ring ($e<0.007$), centered on the star (see Methods). Simulations predict that 
disk eccentricity is no longer excited beyond $\sim3\times$ the planet orbital radius \citep{Bitsch2013, Padgett2026}, thus the circular morphology of the millimeter-dust ring at only 2.5$\times$ the orbital radius cannot be explained by this effect. Alternatively, planets on circular orbits can dampen disk eccentricity \citep{Kley2019, Penzlin2021}. While no planet has been detected between planet 2b and the millimeter dust ring, the scattered-light images \citep{vanCapelleveen2025} show evidence for a shallow gap at $\sim$130 au, with tentative evidence for a positive gradient in the \dvphi{} profile (Figure \ref{fig:radialprofiles}), which could be explained by the presence of an undetected sub-Jovian planet \citep{Byrne2026}. The lack of clear evidence in the \dvphi{} profile may be due to the fact that the gap width is smaller than the spatial resolution. The presence of a third planet would naturally account for both the circularity and the radial location of the millimeter dust ring.

\begin{figure}
    \centering
    \includegraphics[width=\textwidth]{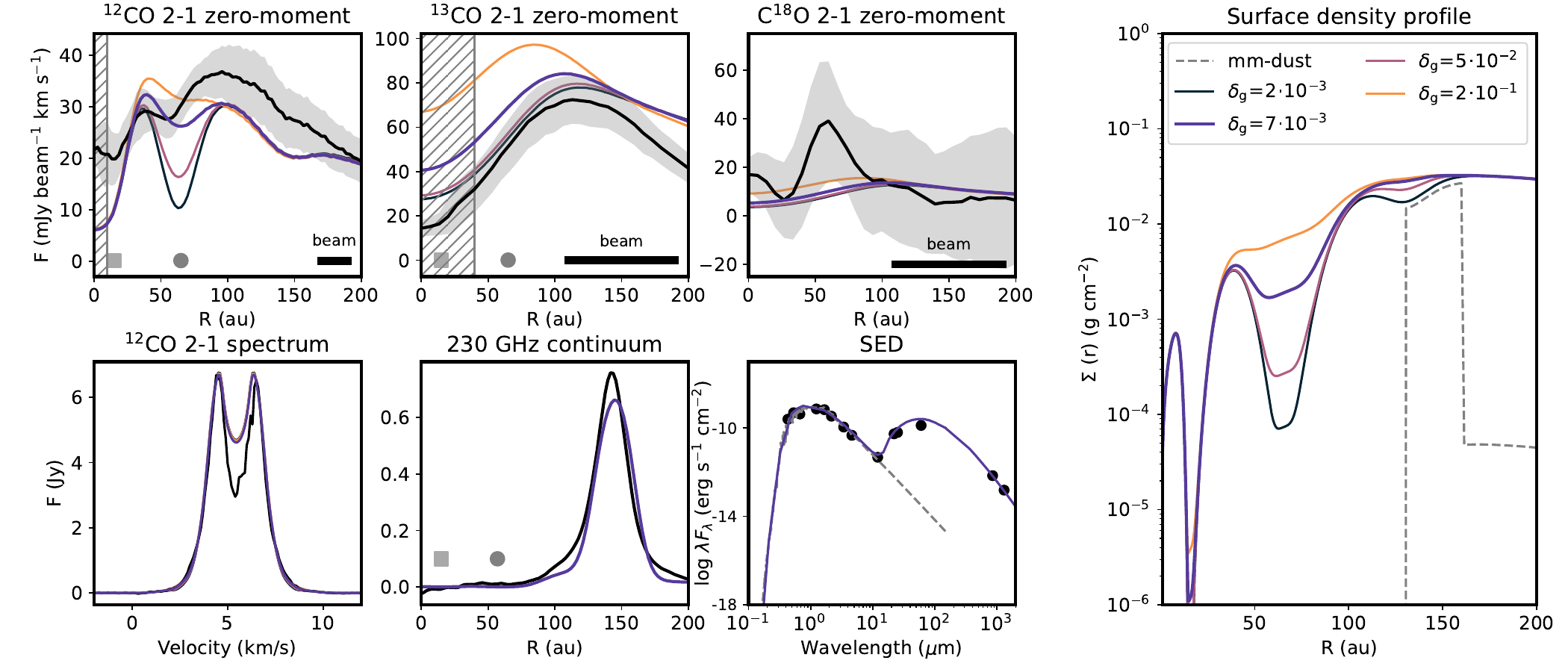}
    \caption{Results of DALI modeling of the gas and dust surface density structure of the \wisa disk. The top three panels show the radial profiles of the beam-convolved zero-moment map of the three CO isotopologues in black, with a grey hatched region indicating the uncertainty based on the standard deviation of the azimuthal average. The colored profiles indicate different models. The bottom three panels show the integrated $^{12}$CO spectrum, the radial profile of the 1.3mm continuum and the SED, with again the data shown in black. The right panel shows the gas surface density profiles of the final model, explored for different values of $\alpha$, corresponding to different values of $\delta_{\rm gas}$ in different colors, which set the depth of the gap. The model with $\delta_{gas}=5\times10^{-2}$ for planet 2b} is most consistent with the data.
    \label{fig:finaldali}
\end{figure}

%
%
To quantify how protoplanets sculpt their surrounding disk, we have estimated the gas surface density profile, in particular the gap profiles, of the \wisa disk using ALMA CO isotopologue data. Whereas CO is the second-most abundant gas molecule in the Universe, converting its emission to a gas surface density in a protoplanetary disk is not trivial, particularly inside a dust gap, as its abundance, optical depth and temperature vary throughout the disk due to various physical processes \citep[e.g.][]{Miotello2023}. Inferring the gas disk structure from line observations requires a thermo-chemical disk model, in which the gas temperature, excitation and abundances are computed self-consistently \citep{Bruderer2013}. We fit a parametric gas surface density profile which includes an exponentially tapered power law for the general disk profile, and a parametrized radially asymmetric gap profile representative of theoretical eccentric gap profiles for each of the planets (Methods). The model output is matched with the Spectral Energy Distribution (SED) from optical to millimeter wavelengths, millimeter-continuum ring and the CO isotopologue emission. Our best-fitting model is presented in Figure \ref{fig:finaldali} and the model parameters are given in the Methods section. In the best-fit model, both gaps are consistent with an eccentric gap profile where the outer gap width is $\sim$1.5-2.5 times wider than the inner gap width. A shallow third gap is added to the model at 130 au to explain the gap in the scattered light image. This gap is undetected in  $^{12}$CO emission, consistent with its optical depth at this radius.

%
%
The derived \wisa gas gap depths are tentatively consistent with the theoretical estimates for a planet gap for the known planet masses and local scale height for moderate $\alpha$-viscosity (see Methods), assuming similar relations of gap depth and viscosity as for circular gaps \citep{Fung2014}, where the known planet masses could break the general degeneracy between planet mass and $\alpha$-viscosity. A third planet is needed to explain the circularity of the disk outside the eccentric orbit of planet 2b. 
The formation of multiple giant planets is consistent with core formation through pebble accretion, because a sufficiently high pebble flux through the disc that allows for the formation of one giant planet automatically allows for that of multiple giant planets \citep{Lambrechts2014,Bitsch2019}. The series of decreasing planet masses with radius in \wisa, similar to what is seen in the Solar System, potentially hints at sequential giant planet formation \citep{Lau2024}, although the wide separations between the planets remains challenging to explain. Considering the presence of at least 14 $M_{\rm Jup}$ of mass locked up in protoplanets and the relatively low gas surface densities in its inner region ($<$150 au), the \wisa disk may have almost finalized its giant planet formation process. Considering that the planet mass is significantly higher compared to the integrated local gas surface density, this would also have slowed down inward migration \citep{KleyNelson2012}, which could explain why these planets remain at such wide orbits. Interestingly, the PDS~70 system which contains two confirmed massive protoplanets of 2-10 $M_{\rm Jup}$ \citep{Wang2021}, appears to have a relatively low gas surface density as well compared to its planet masses \citep{Portilla2023}, indicating a potentially similar scenario. This could imply that planets in more massive disks where gas gaps have been identified \citep{vanderMarel2016-isot} are (much) lower in mass, thus earlier in their planet formation process, and therefore those planets remain undetectable with current facilities.  

%
%
Whereas signatures of planet-disk interaction have been detected in several other protoplanetary disks, either through gaps identified as drops in line emission \citep[e.g.][]{vanderMarel2016-isot,Dong2017,Leemker2022}, pressure gradient variations inferred from the gas rotation curve \citep[e.g.][]{Teague2018,Rosotti_ea_2020,stadler_ea_2025}, or localized deviations from Keplerian motion \citep[e.g][]{Pinte2019,Izquierdo_ea_2023}, none of these signatures have yet been convincingly detected around a confirmed protoplanet. Even in the PDS~70 system, which contains two well-confirmed protoplanets \citep{Keppler2018,Haffert2019}, ALMA CO observations have not detected a localized kinematic signature or resolved two distinct gaps in either the CO intensity maps \citep{Portilla2023} or the pressure gradient \citep{Keppler2019}, due to the planets' proximity to the star and to each other. \wisa thus provides a key and so far unique laboratory for quantifying gap profiles and gas kinematics in the vicinity of planets that shape them, which are essential to understand planet-disk interactions and the formation of giant planets.

\newpage

\clearpage

\setcounter{figure}{0}    

\renewcommand\figurename{Supplementary Fig.}
\renewcommand\tablename{Supplementary Table}%

\section{Methods}\label{sec:methods}

\subsection{Observations}\label{sec:obs}
The ALMA observations of the \wisa disk were taken as part of program 2025.1.00433.S (PI Van Capelleveen) in Band 6 at an angular resolution of $\sim$0$\farcs$12, in two configurations to cover spatial scales out to 5$\farcs$8. The long baseline data were taken on October 23rd, 2025, and the short baseline data on December 18th and 21st, 2025. Details on integration time, baseline ranges, and calibrators are listed in Supplementary Table \ref{tbl:obs}. The data were pipeline-calibrated using CASA version 6.6.6 following the provided calibration scripts. The data were inspected, and no additional flagging was applied.

The spectral settings are set as follows: one spectral window was centered on $^{12}$CO 2--1 (230.538 GHz) with a bandwidth of 117 MHz and a channel width of 70.6 kHz (0.08 km s$^{-1}$) and two spectral windows were centered on 217 and 232.5 GHz with a bandwidth of 1875 MHz and a channel width of 1.1 MHz (continuum windows). In the long baseline data, the final spectral window was incorrectly set on 219 GHz with a bandwidth of 937.5 MHz, missing the CO 2--1 isotopologues. For the long baseline data, this was corrected to one spectral window centered on the $^{13}$CO 2--1 line (220.39868 GHz) and the C$^{18}$O 2--1 line (219.56035 GHz), each with a bandwidth of 469 MHz and a channel width of 488 kHz (0.66 km s$^{-1}$). The continuum was constructed by averaging all spectral windows except the one including the $^{12}$CO 2--1 line, since the line is relatively strong. The central frequency of the continuum is 224.7 GHz ($\lambda$=1.34mm).

The pipeline calibration process includes automatic self-calibration, which improved the SNR on the continuum peak from 40 to 56 for the long baseline dataset and from 60 to 148 for the short baseline dataset when inspecting the product fits files. Additional manual self-calibration was attempted for both datasets, but did not further improve the SNR. Both the continuum and the $^{12}$CO spectral window of both datasets were concatenated. For the $^{13}$CO and C$^{18}$O lines, the low-resolution data were imaged as no high-resolution data exist. 

Continuum imaging was performed using both natural and Briggs weighting with \texttt{robust=0.5}. The deconvolver parameter was set to \texttt{multiscale} with \texttt{scales=[0,5,10,20]} with a pixel size of about 1/5th of the beam size. The mask was drawn with automasking, and cleaning was continued until the emission inside the mask was below 3$\sigma$. Natural weighting resulted in a 0$\farcs$20$\times$0$\farcs$10 beam (PA=-67$^{\circ}$), with a total flux of 70 mJy, $\sigma$=0.018 mJy beam$^{-1}$ and a peak SNR of 83. Briggs weighting with \texttt{robust=0.5} resulted in a 0$\farcs$13$\times$0$\farcs$06 beam (PA=-67$^{\circ}$), with a total flux of 66 mJy, $\sigma$=0.014 mJy beam$^{-1}$ and a peak SNR of 52. Considering the limited flux loss, the latter is chosen for the main continuum image, which is presented in Figure \ref{fig:mainimages}.

Line imaging was performed using natural weighting with the same deconvolver setup as for the continuum. During the cleaning process, a Keplerian mask was used, with parameters fine-tuned informed by the kinematic analysis (see below). In the final images, we assume a stellar mass $M_*$=1.37 $M_{\odot}$, inclination $i$=46.0$^{\circ}$, position angle PA=178.7$^{\circ}$, source velocity $v_{\rm LSR}$=5.45 km s$^{-1}$ and an outer radius of 3$\farcs$0. The $^{12}$CO 2--1 cube was cleaned at 100 m s$^{-1}$ velocity resolution. The resulting cube has a beam size of 0$\farcs$19$\times$0$\farcs$10 and a rms of 2.5 mJy beam$^{-1}$ based on line-free channels. The $^{13}$CO 2--1 and C$^{18}$O 2--1 cube were cleaned at the channel resolution of 0.66 km s$^{-1}$ with the same procedure, resulting in a beam size of 0$\farcs$66$\times$0$\farcs$53 and a rms of 2.2 mJy beam$^{-1}$. The C$^{18}$O 2--1 emission was not detected, even after cleaning at 1.5 km s$^{-1}$ resolution. 

The line cubes were integrated to produce intensity moment maps using \texttt{bettermoments} \citep{Teague2018bm}, with the same Keplerian mask used for cleaning. In addition, continuum images and $^{12}$CO 2--1 line cubes were generated with a circular beam with a beam size of 0$\farcs$15, through a combination of \texttt{uvtaper} and \texttt{restoringbeam} with Briggs weighting and \texttt{robust=0.5}. For certain purposes in our analysis (see Section  \ref{sec:kins}), a circular beam is preferred \citep{loomis_ea_2025}. Due to the strong sampling in V compared to U, it was not possible to generate a circular beam with uvtaper alone. Since the total flux was recovered in this image, it is considered a reasonable imaging product.

\subsection{Continuum emission}
\label{sec:visibility}
In order to analyze the morphology of the continuum emission, we use visibility modeling with a parametric ring model of the radial intensity profile, using \texttt{galario} \citep{Tazzari2019}, following previous work \citep[e.g.][]{Pinilla2018tds}. We assume that the ring has an asymmetric Gaussian shape, with a different width on either side of the peak. This means that the radial intensity profile is described as a function of radius $r$ (in arcsec) as follows:
\begin{equation}
I(r)=\\
\begin{cases}
    I_1 \exp\left(-\frac{(r-r_{c1})^2}{2 r_{w1a}^2}\right) & \text{for r$<r_{c1}$} \\
    I_1 \exp\left(-\frac{(r-r_{c1})^2}{2 r_{w1b}^2}\right) & \text{for r$>r_{c1}$} \\    
\end{cases}
\end{equation}

with prefactor $I_1$, peak $r_{c1}$ and widths $r_{w1a}$ and $r_{w1b}$ for inner and outer width, respectively. In addition to the Gaussian parameters, we fit the inclination $i$ and position angle PA defined as the angle towards the major disk axis east-of-north, and the central position of the ring with dRA and dDec. We run the MCMC fit with broad priors on all free parameters, using 80 walkers, 3000 steps, and a burn-in of 1000 steps. Convergence is found for the best-fit parameters with uncertainties listed in the left column (Single ring) of Supplementary Table \ref{tbl:galariofit}, the visibility model and data shown in blue in Supplementary Figure \ref{fig:vismodel}. Although the model converges, the visibility plot demonstrates that some flux is missing at the shortest baselines, i.e., some larger spatial scale emission is missing. Therefore, motivated by \citep{Sierra2025}, we perform a second fitting procedure using a \texttt{galario} fit of the sum of two asymmetric Gaussian rings, i.e.

\begin{equation}
I(r)=
\begin{cases}
    I_1 \exp\left(-\frac{(r-r_{c1})^2}{2 r_{w1a}^2}\right)  & \text{for r$<r_{c1}$} \\
    I_1 \exp\left(-\frac{(r-r_{c1})^2}{2 r_{w1b}^2}\right) & \text{for r$>r_{c1}$} \\ 
\end{cases}\\
+ \\
\begin{cases}
    I_2 \exp\left(-\frac{(r-r_{c2})^2}{2 r_{w2a}^2}\right) & \text{for r$<r_{c2}$} \\
    I_2 \exp\left(-\frac{(r-r_{c2})^2}{2 r_{w2b}^2}\right) & \text{for r$>r_{c2}$} \\ 
\end{cases}
\end{equation}

We run the MCMC fit using 120 walkers, 5000 steps, and a burn-in of 1000 steps. Convergence is found for the best-fit parameters with uncertainties listed in the right column (Double ring) of Supplementary Table \ref{tbl:galariofit}, the visibility model and data shown in red in Supplementary Figure \ref{fig:vismodel}.

The double-ring model shows better agreement with the visibility data, fitting the flux even at the shortest baselines. The geometric fit parameters (inclination, position angle, and center) are almost identical to the parameters found for the single ring model. The first ring has the same peak radius of 1$\farcs$09 but is slightly narrower (0$\farcs$14 or 19 au), whereas the second ring has converged to a highly asymmetric Gaussian at 0$\farcs$63 (84 au radius), dropping off steeply on the inner side and more gradually on the other side (out to 1$\farcs$3, i.e., beyond the first ring). This profile (see right panel in Supplementary Figure \ref{fig:vismodel}) resembles a faint pedestal as seen in other transition disks \citep{Keppler2019,SHuang2024}, although we notice that it is also close to IR ring 2 in the SPHERE image just outside the orbit of \wisb \citep{vanCapelleveen2025}. The offset from the phase center, set to the stellar position, is 0$\farcs$008, well within the ALMA astrometric uncertainty. This means that the maximum eccentricity of the dust ring is at most $<0.007$.

In order to assess the quality of the fit in the image plane, we map the model visibilities of the best fit of the double ring model onto the observed visibilities, subtract these in the visibility plane, and clean both the model and the residuals using the same parameters as for the data. The result is shown in Supplementary Figure \ref{fig:imagevismodels}.

The model image clearly resembles the morphology of the data image, although the emission in the ansae is 25\% fainter. Along the minor axis, the difference is less than 10\%. In the residual image, the maximum residual is seen at the ansae (24$\sigma$), whereas the emission along the rest of the ring is generally below 3$\sigma$, with $\sigma$ the rms level. 

In the residual continuum image of the visibility modeling from Supplementary Figure \ref{fig:imagevismodels}, we search for evidence for a circumplanetary disk (CPD) at both planet locations. No point sources inside the cavity are identified, and there is no significant emission at the planet locations (see rightmost panel in Supplementary Figure \ref{fig:imagevismodels}: planets are indicated with white square and circle, as before). The rms level inside the cavity is 25 $\mu$Jy beam$^{-1}$, setting a 3$\sigma$ upper limit of 75 $\mu$Jy on the CPD flux. We note that the detection of the CPD around PDS70c in ALMA Band 6 observations was only 60 $\mu$Jy with a sensitivity of 15 $\mu$Jy \citep{Fasano2025}, i.e. a factor of 2 deeper. Deeper ALMA Band 7 observations of WISPIT-2 at 15 $\mu$Jy rms \citep{Facchini2026} did not reveal the CPD. 

\subsection{Kinematics}\label{sec:kins}
We modeled and analyzed the \twCOfull{} circular-beam cube using code package \texttt{discminer} \cite{Izquierdo2021}, which fits the channel-by-channel emission of the data assuming a Keplerian and smooth disc. It maximizes the log-likelihood function between the observed and model intensities at each spatial pixel and velocity channel. The best-fit model parameters are then obtained from the posterior distributions as mapped by the MCMC ensemble sampler \verb|emcee| \cite{Emcee_2019}. A double power law was prescribed for the model peak intensity profile to account for the observed gas cavity (see Supplementary Figure~\ref{fig:vel_curve}a) and incorporates both upper and lower surface emission contributions in the channel maps.

We adopted initial guesses for the model parameters based on the inclination, position angle, and stellar mass as reported by \cite{vanCapelleveen2025}. The remaining parameters were initialized using a prototype model that yields channels similar to those in the data. For the 25 model parameters, we employed 256 walkers and ran the fit for $24,000$ steps, continuing until the marginalized posteriors converged to Gaussian distributions. An overview of the best-fit model parameters is provided in Supplementary Table~\ref{table:attributes_parameters}.

In Supplementary Figure~\ref{fig:dm_channels}, we show a gallery of selected \twCO{} channel maps to demonstrate the goodness of our \texttt{discminer} fit to the data. The lower surface is clearly visible in the emission morphology due to the system's relatively high inclination and is well-captured by our model. Next, we fit a Gaussian to the local line profile of each pixel to obtain peak-intensity and centroid-velocity moment maps for the data and model cubes. Looking at the line-of-sight velocity residuals (Supplementary Figure~\ref{fig:kinematic_detection}), we note that the disk is dynamically quiescent beyond the continuum and IR ring 2 (outermost black ellipse). Inside the continuum cavity, the residuals are dominated by alternating blue and red-shifted patterns mirrored around the disk minor axis. These patterns signal super- and sub-Keplerian rotation, indicative of pressure bumps and gaps, respectively, and illustrations of these patterns are provided in \cite{Izquierdo_ea_2026b}.

From the line-of-sight velocity $\upsilon_0$, we derive radial profiles of the azimuthally averaged rotational velocity $\upsilon_\phi$, for both the data and the model, using the analytical method implemented in the \texttt{discminer} framework:
\begin{equation} \label{eq:vphi_dm}
    \upsilon_\phi = \frac{\psi}{4\sin{\frac{\psi}{4}}\sin{|i|}} \left<\left|\upsilon_{0} - \upsilon_{{\rm LSRK}}\right| \right>_{\psi},
\end{equation}
where $i$ is the disk inclination, $\upsilon_{\rm LSRK}$ the disk's systemic velocity, and $\psi$ denotes the angular extent of the azimuthal section where the averages are computed \citep[see Appendix C in][for derivation]{Izquierdo_ea_2023}. Analogous to other studies, the rotational velocity is assumed to be azimuthally symmetric and dominant over the radial and vertical velocity components.
To accurately recover the centroid velocity map of the upper surface, we fit a two-component double-Bell function to the line profile in each pixel \citep{izquierdo_ea_2025}. This decomposition is required here to disentangle the contributions from the upper and lower surfaces, which manifest as a double-peaked structure in the local line profiles. Following \cite{stadler_ea_2025}, we extract the velocity radial profiles starting one beam width from the disk center in steps of a quarter of the beam width (5\,au). The uncertainties of each bin are reported as the standard deviation divided by the square root of independent beams along each radial annulus. Deviations from Keplerian rotation can then directly be obtained via $\delta\upsilon_\phi=\upsilon_{\phi, \rm{data}}-\upsilon_{\phi, \rm{model}}$. Because the model is convolved with the observational beam, subtracting the model $\upsilon_{\phi, \rm{model}}$ instead of the ideal Keplerian profile partially accounts for beam-smearing effects near the disk center, which would otherwise produce artificial sub-Keplerian rotation \citep[e.g.,][]{Keppler2019, Pezzotta_ea_2025}.

Supplementary Figure~\ref{fig:vel_curve}~(c)~and~(d) show the radial profiles for the azimuthally averaged rotational velocity \vphi{} and its deviations from Keplerian rotation \dvphi{}, respectively. The disk is not strongly dynamically perturbed beyond the dust continuum peak, as already seen in the velocity residual map. The dust continuum itself aligns with a radially decreasing \dvphi{}-profile, indicative of a strong gas pressure bump at this location, as demonstrated by \cite{Teague2018, Rosotti_ea_2020, stadler_ea_2025}. Consistent with the aforementioned observations, the continuum ring is observed at a closer-in radius, rather than at the location where the negative \dvphi{}-gradient reaches half of its maximum value. This observation does not align with theoretical expectations, as the continuum ring is expected to be located at the center of a pressure bump \citep[as discussed in][]{Rosotti_ea_2020}. Mechanisms that may explain the observed offsets are velocity asymmetries, including spiral structures and vertical motions, which can systematically bias the measured \dvphi{} \citep[][Figure D1]{stadler_ea_2025}. Additionally, beam smearing effects arising from strong line intensity gradients may also contribute \citep[e.g.,][]{Pezzotta_ea_2025}. Still, real vertical variations in the pressure structure can cause shifts in the radial positions of \dvphi{}-gradients at higher disk layers relative to the midplane dust ring locations, as suggested by observations of HD\,163296 \citep{Izquierdo_ea_2023}. Comprehensive multi-tracer observations are required to reconstruct the disk's three-dimensional pressure structure accurately.

On the contrary, the IR gap 3, in which \wisb resides, is shaped by a strong positive \dvphi{}-gradient, indicative of a pressure gap. The peak brightness temperature also shows a clear deficit (Supplementary Figure~\ref{fig:vel_curve}b). Hence, this pressure feature is likely not driven by temperature but rather by density variations \citep{Rab_ea_2020, Izquierdo_ea_2023}, suggesting a real gas-surface-density gap at the planet's location. Interestingly, the pressure gap feature appears to extend outward more radially than the planet's location. From theory, we would expect the planet to reside at the center of a positive \dvphi{}-gradient \citep[e.g.,][]{Teague2018}. However, if the planet is on an eccentric rather than circular orbit, the gap becomes asymmetric and more extended outward, as similarly observed in the integrated intensity radial profiles (see Section below on gap profiles). 

Regarding eccentric motions in the disk, to first order, we would expect an eccentricity-squared modification on the averaged rotation curve. A mildly eccentric disk of $e \approx 0.1$ (similar to \wisb's orbit) would modify the azimuthally averaged rotation curve only at the $\sim1\%$ level.
However, observed \dvphi{}-deviations near \wisb and the concentric dust continuum ring reach several percent of the Keplerian speeds (see Figure \ref{fig:radialprofiles}). Thus, the observed pressure modulations induced by the planets and the pressure bump are several times greater than the maximum effect expected from disk eccentricity.

Radially decreasing \dvphi{} points are also observed co-located at IR rings 3 and 2. At Ring 2, the \dvphi{}-profile exhibits a tentatively increasing gradient that may indicate the presence of a very shallow gas density gap feature. This observation could account for the detection of this ring exclusively in IR and not in the continuum. Lastly, beyond IR ring 2 at $R\sim110-120\,\rm{au}$, the \dvphi{}-gradient increases once more, indicative of a pressure gap that also closely coincides with the IR gap detected there. 

Investigating disk kinematics is valuable as it enables the detection of embedded protoplanets via their imprints on the gas dynamics within protoplanetary disks \citep{Perez2018, Pinte_ea_PPVII}. Surface density gaps, potentially carved out by massive planets, generate azimuthal velocity flows due to the gas pressure gradient \citep{stadler_ea_2025}. These azimuthally symmetric features dominate our 2D velocity residual map inside the continuum ring (Figure~\ref{fig:kinematic_detection}), making it hard to detect a superimposed localized planetary signal within it. To circumvent this problem, one can additively fold the centroid residual map along the disk minor axis, thus canceling out these azimuthally symmetric rotational contributions \citep{Izquierdo_ea_2026b}. This folding is achieved by adding the line centroid velocities of one half of the disk from those of the other half, split along the disk's minor-axis symmetry ($\phi=\pm90\degree$).

To search for strong deviations from Keplerian motion, we extract peak velocity residuals from the folded map in radial annuli sampled every fifth of the beam size. The global peak value is identified with a significance of $10\,\sigma_{\rm MAD}=0.38\,\mathrm{km\,s^{-1}}$ at a radius of $46\,$au and azimuth of $38\degree$ in the disc frame ($R=$0$\farcs$309, PA$=207\degree$ on the sky, which is $0\farcs19$ apart from the current \wisb location) as can be seen in Supplementary Figure\,\ref{fig:folded_map_peaks}. The measure ${\sigma_{\rm MAD}=1.4826\times \mathrm{median}(|p-\mathrm{median}(p)|)}$ provides an estimate of the statistical dispersion of the peak residuals (p), based on the median absolute deviation (MAD), equivalent to the standard deviation for normally distributed data while remaining robust to a small number of outliers such as those expected from planet-driven perturbations \citep[see][for details]{Izquierdo_ea_2026b}.

The peak velocity residuals are subsequently grouped into eight radial and eight azimuthal clusters identified via a K-means algorithm \citep{Izquierdo2021}, forming a two-dimensional segmentation in ($r, \theta$)-space, within a radial range extending beyond 1.5 major beam sizes from the disk center, out to $350\,$au. Within each cluster, the distribution of peak residuals is characterized by its spread and statistical significance relative to all other clusters. Comparisons between each identified ($r,\theta$)-cluster and the global background are performed to evaluate whether the amplitude of the signal and its spread in a given region are anomalous. This framework provides a quantitative measure of how strongly localized velocity perturbations deviate from the system's typical kinematic dispersion.

Clusters of residual points that exceed the predefined statistical threshold above $3\sigma$ from the background residual points are classified as significantly perturbed regions. The radial and azimuthal coordinates of these K-means clusters thereby constrain the spatial location of localized kinematic disturbances within the disk. We identified three localized significant clusters: one in radius at 54\,au with $25\sigma$ and two in azimuth at $37\degree$ and $53\degree$ with $3.7\sigma$ and $7.8\sigma$ significance, respectively. The azimuthal K-means cluster statistics is shown in Supplementary Figure~\ref{fig:phi_clusters}. We note that even though there are a few peak residual points above $3\,\sigma_{\rm MAD}$ at an azimuthal location of $\sim-70^{\circ}$, the variance of their associated cluster is not significant ($1.5\times10^{-3}\,\mathrm{km^2/s^2}\approx2\sigma$). Since this cluster is at \wisb's radial location, it could be related to asymmetric flows induced by the planet around its orbit (e.g., spiral signatures and meridional circulation). Similar azimuthally offset clusters of strong residuals are reported from planet-disk interaction simulations in \cite[][Fig.~3]{Izquierdo_ea_2026b}. 

The weighted mean center of the significant clusters is located at $R=54\pm11\,$au and $\phi=48\pm5\degree$ in the disc frame reference. The reported uncertainties correspond to the standard deviations of the locations of the peak residuals within the detected clusters. For a visualization, the left panel of Suppl.~Fig.~\ref{fig:phi_clusters} shows the clusters' similar azimuthal widths in the yellow contours. For both detected azimuthal clusters, the azimuthal center is calculated by weighting each point according to its normalized individual significance (${\phi_\mathrm{cent}=(\sum_{i}^{N=2} \phi_i\times\sigma_i)/(\sigma_1+\sigma_2)}$). We also ran the algorithm using a partitioning with 8 to 12 ($r, \theta$)-clusters. In all cases, the retrieved central cluster locations were consistent within the error bars. 

The weighted cluster centers correspond to ${R=342^{+57}_{-61}\,\mathrm{mas}}$ and ${\mathrm{PA}=216\degree\pm5\degree}$ projected to sky coordinates. For comparison, the current on-sky position of \wisb is  $R=316\pm2\,\mathrm{mas}$, PA$=241.9\degree\pm0.2\degree$ (see Section on orbital analysis below). While the radial locations of the cluster center and planet agree within $1\,\sigma$ of the reported uncertainties, the position angles are offset by $26\degree{}$, or one beam size (separation $\approx 0\farcs15$), ahead of the disk's clockwise rotation.
Several factors may account for this azimuthal offset. First, \wisc is located within a few beam sizes of b, and the extent to which such a massive planet \citep[$\sim10M_\mathrm{Jup}$,][]{Lawlor2026} perturbs the gas dynamics near b remains unclear. Such strongly perturbed motions are amplified by the possibility that both planets are on eccentric orbits. The influence of orbital eccentricity on gas kinematics in the vicinity of planetary orbits has not yet been thoroughly investigated. Second, b is positioned very close to the disk minor axis. As a result, the planet's azimuthal velocity perturbations are barely detectable due to the component's orthogonality in the line-of-sight projection. Consequently, the azimuthal component of the planet's (trailing) spiral wake closest to the minor axis is suppressed. For this reason, the clustering algorithm is also less effective for planets located near the disk's minor axis \citep[Fig.~10]{Izquierdo2021}, along which the residual map must be folded. Taken together, these effects may account for the observed azimuthal offset.

In conclusion, our findings demonstrate that localized deviations from Keplerian rotation, identified through additive folding and clustering analyses, are directly associated with the presence of an embedded protoplanet. This outcome offers long-sought confirmation of the gas-kinematic search for embedded planets using molecular line observations.

\subsection{Re-analysis of the SPHERE image}
\label{sec:SPHERE}
We extend the geometric analysis of the $H$-band ($1.5$--$1.8,\mu$m) scattered light VLT/SPHERE image of \wisa previously presented by \citep{vanCapelleveen2025, Byrne2026}. The methods used to characterize the disk geometry are described therein. We fix the inclination $i$ and position angle PA to the values retrieved from the CO kinematics fitting ($i$=46$^{\circ}$ and PA=178.7$^{\circ}$,  defined as the angle of the major axis measured east-of-north). 

We perform ellipse fitting on a deprojection of the $H$-band image to determine the intrinsic eccentricities of the structures. The deprojection was performed using the height profile obtained in \cite{vanCapelleveen2025, Byrne2026} of $h = 0.0026r^{1.79}$ au. 
The resulting fits to the deprojected image are shown in the right panel of Supplementary Fig. \ref{fig:ellipsefits} and parameters in Supplementary Table~\ref{tab:constrained_summary}. Ellipse fits were not performed for Ring~1, Gap~1, or Ring~0 due to insufficient signal. 

Rings~2 and 3 are found to be weakly eccentric, with $e = 0.132 \pm 0.013$ and $e = 0.128 \pm 0.070$, respectively. The separation of Gap~2 was not previously estimated from the gap itself, and we obtain a separation of $131.166 \pm 0.889$~au from the minimum of the inverted Gaussian fit, corresponding to the midpoint of the gap. 

\subsection{Orbital analysis protoplanets}
\label{sec:orbits}
Using the new inclination, position angle, and stellar mass from the kinematic modeling of the outer disk in the CO data (Supplementary Table \ref{table:attributes_parameters}), we re-fit the orbits of both protoplanets (Supplementary Figure \ref{fig:WISPIT_2b_orbit}). Our orbital fitting was performed using the orbit fitting package \texttt{orbitize!} \citep{Blunt2020}, sampling posterior distributions with a parallel-tempered affine-invariant Markov Chain Monte Carlo (MCMC) sampler implemented via \texttt{ptemcee} \citep{Vousden2016}. We follow the methodology of \citep{vanCapelleveen2025}. 

For \wisb, we restrict the semi-major axis to lie between 40 and 100 au, and impose an additional physical constraint to prevent the orbit from crossing the disk by requiring the apastron to satisfy $Q = a(1+e)\leq 70$ au, where $Q$ is apastron, $a$ is semi-major axis, and $e$is  the eccentricity. This adds to the methods presented in \citep{vanCapelleveen2025} by ensuring that highly eccentric orbits do not evolve in a way that would physically perturb the disk. We include the astrometry from \citep{vanCapelleveen2025} and \citep{Vach2026}, along with a new $H$-band measurement obtained on UTC date 2025-09-24 (program ID 115.29HG.002, PI Ginski), described in \citep{GRAVITY2026}. The astrometry is extracted following the methods of \citep{stolker2020}, applying simplex minimization to the ADI/PCA reduction with 7 principal components; equivalent to the reduction in \citep{GRAVITY2026}. We used 200 walkers and 10000 steps to sample the posterior, and removed 20\% as burn-in. In addition to the included systematic errors of the method described in \citep{vanCapelleveen2025,GRAVITY2026}, uncertainty in plate scale, true north correction, and the centering precision of 2.5\,mas of the star behind the coronagraph, as described in the SPHERE manual (\href{https://www.eso.org/sci/facilities/paranal/instruments/sphere/doc.html}{https://www.eso.org/sci/facilities/paranal/instruments/sphere/doc.html}), were included in the error budget. This yielded a separation of $315.8 \pm 2.1$\,mas and a position angle of $241.86 \pm 0.20^\circ$. 
This additional epoch extends the observational baseline to $\sim$ 23 months, compared to 18 months before \citep{vanCapelleveen2025}. While still relatively short, it provides key additional orbital information. The resulting posterior distributions yield an updated semi-major axis of 64.40$^{+2.95}_{-4.47}$ au. The inferred eccentricity is 
0.12$^{+0.04}_{-0.06}$, indicating 
that the planet orbit is not circular, as suggested by the disk observations. The combined eccentric orbits result in a time-averaged mean orbital separation of $r_p=64.7\pm3.6$ au.

For \wisc, we use the astrometry presented in \citep{Lawlor2026}. Although the observational baseline is short ($\sim$ 6 months), we detect significant orbital motion of approximately 2$^{\circ}$. Assuming the planet remains confined within the cavity in which it resides, we impose a maximum apoastron constraint of  $Q\leq 25$ au. The resulting posterior distributions are less well constrained than for \wisb, as expected given the limited baseline, but indicate that \wisc shares the same direction of motion. The inferred semi-major axis is 13.66$^{+2.37}_{-2.35}$ au. The eccentricity is potentially higher than that of \wisb, with a value of 0.29$^{+0.28}_{-0.20}$. The combined eccentric orbits result in a time-averaged mean orbital separation of $r_p=14.5\pm1.8$ au.

The analysis of the planet orbits indicates that they are consistent with moderate eccentricity, which in turn is consistent with the estimates of the disk rings in the scattered light image and the CO gaps seen in the ALMA observations.

\subsection{Theoretical considerations on the gap profiles}\label{sec:gaps}
Planets can carve gaps in the gas disk under appropriate conditions \citep{Paardekooper2023}. We have chosen to parametrize the gap profiles using physical prescriptions from the literature in order to quantify their properties.

For massive planets with planet-to-star mass ratio $q>10^{-4}$ on circular orbits, prescriptions for gas gaps have been synthesized from hydrodynamic simulations in disks with $\alpha$=10$^{-1}$--$10^{-3}$ and $h$=0.04--0.1 \citep{Fung2014, Fung2016}. For planets with $10^{-4}<q<5\times10^{-3}$, the gap depth $\delta_{gas}=\Sigma_{gap}/\Sigma_0$ is: 
\begin{equation}
\label{eq:fung1}
    \delta_{gas} = 0.14\left(\frac{q}{10^{-3}}\right)^{-2.16}\left(\frac{\alpha}{10^{-2}}\right)^{1.41}\left(\frac{h/r}{0.05}\right)^{6.61},
\end{equation}
and for high-mass planets with $5\times10^{-3}<q<10^{-2}$,
\begin{equation}
\label{eq:fung2}
    \delta_{gas} = 4.7\times10^{-3}\left(\frac{q}{5\times10^{-3}}\right)^{-1.00}\left(\frac{\alpha}{10^{-2}}\right)^{1.26}\left(\frac{h/r}{0.05}\right)^{6.12},
\end{equation}
where $\Sigma_{gap}$ is the gas surface density inside the gap and $\Sigma_0$ is the initial surface density prior to gap opening. The width of these gaps is typically 5$R_H$, where $R_H$ is the Hill radius of the planet, which is given by $R_H=r_p (q/3)^{\frac{1}{3}}$.

However, this relation is only valid for planets fixed at circular orbits. Simulations have shown that planet orbits quickly become eccentric for $q\gtrsim3\times10^{-3}$ \citep{KleyDirksen2006,Teyssannier2017}, where in an azimuthally averaged profile the outer width of the gap becomes wider and the gap shallower. The $^{12}$CO gap profiles indeed show such evidence in both the integrated intensity and the $\delta v_{\phi}$ profiles (Figure \ref{fig:radialprofiles}): the outer edge of the gap opened by planet 2c at 15 au extends out to $\sim$40 au, and the outer edge of the gap opened by planet 2b at 65 au extends out to $\sim$100 au, which means that for both gaps the gap's outer half-width, $w_{\rm half,out}$, is at least two times wider than the gap's inner half-width, $w_{\rm half,in}$. 

Since the detected planets in the \wisa system have planet-to-star mass ratios of 6.4$\times10^{-3}$ and 3.6$\times10^{-3}$ (computed using the dynamical stellar mass of 1.37 $M_{\odot}$ derived above), this implies that their orbits should have become eccentric. Re-analysis of the planet orbits and IR rings  have confirmed that this is indeed the case. 

Unfortunately, detailed prescriptions for depth and width of eccentric gap profiles carved by massive planets are not available in the literature. Parameter explorations are generally limited to planet mass or eccentricity only \citep{KleyDirksen2006,Hosseinbor2007,Bitsch2013,Teyssannier2017,Muley2019}, rather than a full parameter grid. However, these studies all show similar qualitative behavior for gaps opened by moderately eccentric planets ($e \sim 0.1$–0.2), such as found here: the radial gap profile becomes asymmetric, with the gap's outer half-width typically $\sim$2–3 times larger than in the circular-orbit case, and the gap depth is shallower by roughly a factor of two. Therefore, the gas gap profiles of such gaps can be reasonably parametrized with the following equation:

\begin{equation}
\label{eqn:widegap}
    G(r)=1-\frac{1-2\delta_{gas}}{1+\left(\frac{r-r_p}{w(r)}\right)^4}
\end{equation}
where $\delta_{\rm gas}$ is taken from eqn. \ref{eq:fung1} or \ref{eq:fung2}. 

In eccentric gap regimes, the gap depth $\delta_{\rm gas}$ may be regulated by complex edge dynamics rather than a simple planet–$\alpha$-viscosity torque balance and consequently, it need not perfectly follow standard circular scaling. However, for moderately eccentric planets it has been shown that $\alpha$ continues to regulate the gap depth  \citep[][Eqn. 30]{Teyssannier2017,SanchezSalcedo2023}. We adopt this framework as a reasonable first-order estimate to provide a physical connection between the confirmed planets and the newly observed gas gaps.

The half-width $w_{\rm half}(r)$ is defined as
\begin{equation}
\label{eqn:gapwidth}
w_{\rm half}(r) =
\begin{cases}
w_{\rm half,in} & r < r_p \\
w_{\rm half,out} & r \ge r_p
\end{cases}
\end{equation}
The power in the nominator is chosen to have a value of 4 to create a relatively steep edge profile, rather than a gradual Gaussian profile. The final surface density profile is then defined by the background profile, set by the general disk evolution as the result of accretion \citep{Lesur2023}, multiplied with $G(r)$ for each planet in the system. For lower mass planets $<$1 $M_{\rm Jup}$ on circular orbits, prescriptions of the gap depth and width have been well established \citep{Kanagawa2015,Kanagawa2016}. 

\subsection{Thermal-chemical modeling}\label{sec:dali}
By quantifying the gas gap profiles with the CO isotopologue data, predictions from planet-disk interaction models can be tested. Since the planet masses are known and the gas scale height can be constrained with the observations, the gap depth likely depends primarily on the $\alpha$-viscosity (see discussion above). However, the CO isotopologue emission of a protoplanetary disk cannot be converted directly into a gas surface density, as its intensity depends on a large number of properties that vary strongly throughout the disk. These properties include optical depth, local abundance, freeze-out, photodissociation, decoupled gas/dust temperature and excitation, which depend on local disk conditions such as the gas-to-dust ratio, vertical structure, radiation field, shielding, and chemistry \citep[e.g.][and references therein]{Bruderer2012}. In order to characterize the gas surface density profile of the gas disk around \wisa, we use a thermal-chemical modeling code called Dust And LInes (DALI) \citep{Bruderer2013}, which has been designed specifically for the purpose of interpreting spatially resolved CO isotopologue emission in disks, in particular inner gas gaps in disks \citep{Bruderer2014,vanderMarel2016-isot,vanderMarel2018,Leemker2022}. The input of DALI is a parametric disk density distribution of dust and gas in radial and vertical directions and an input stellar radiation field. The final structure is ray-traced at the relevant wavelengths to construct the SED and the three observed CO 2--1 isotopologue line cubes at the derived inclination. The resulting model cubes are integrated and convolved with the same beam as the observations. 

Thermal-chemical modeling is computationally expensive and contains many free parameters and assumptions, so a modeling procedure to obtain a best fit is not feasible, and the retrieved model structure cannot be considered unique. However, as in previous work, this type of modeling can be used to impose reasonable constraints on the overall gas and dust structure and its gap properties, which can be compared with gaps expected from planet-disk interaction models. By simultaneously matching the Spectral Energy Distribution, the millimeter-dust ring, and the three CO 2--1 isotopologue emission profiles, the derived gas and dust density structure can be considered representative. 

The DALI model is set up with a radial range from the sublimation radius $r_{\rm sub}$ to the outer radius $r_{\rm out}$. The grid consists of 300 cells in the radial direction and 60 cells in the vertical direction, with the radial grid logarithmically spaced to capture the strong temperature gradients at both the inner and cavity edges of the disk. The vertical structure (gas scale height described as $h(r)=h_c(r/r_c)^{\psi}$), including the dust settling as described with the settling parameter $\chi$, is set up following the standard conventions \citep{Bruderer2013}. 

In the first step, we derive the general outer disk properties by quantifying the background surface density profile of the disk, finding a reasonable match with the SED, millimeter-flux, C$^{18}$O non-detection, and global $^{12}$CO and $^{13}$CO intensity curves. For the stellar radiation field, we use a blackbody curve with previously derived stellar properties \citep{vanCapelleveen2025}. For the gas surface density profile, we use an exponentially tapered power-law with two positive power indices rather than the conventional viscous disk model solution \citep{LyndenPringle1974}, in order to match the very deep drop in gas surface density inside the cavity. The power-law is parametrized as:
\begin{equation}
\label{eqn:brokenlaw}
    \Sigma(r)=\Sigma_t\left(\frac{r}{r_t}\right)^{t} {\rm exp}\left(-\frac{r}{r_t}\right)^t
\end{equation}
and shows an initially increasing and then decreasing profile, with a turnover around radius $r_t$. Such a profile is morphologically similar to the solutions of a magneto-hydrodynamic (MHD) wind model \citep{BaiStone2013,Ogihara2015} where accretion is driven by angular momentum removal through magnetically launched disk winds. This mechanism has been previously evoked to explain the deep gas cavities seen in transition disks while maintaining significant accretion onto the star \citep{Martel2022}. Whereas the stellar accretion rate of \wisa is not quantified, the measured H$\alpha$ signal of accretion onto planet 2b \citep{Close2025} provides clear evidence for ongoing inward transport of material in the disk. 

The parameters in Eqn \ref{eqn:brokenlaw} are adjusted to match the global intensity profiles across the disk. The turnover radius $r_t$ is chosen as 150 au, considering the general shape of the $^{13}$CO emission profile, and $t$=1.5 to get a sufficiently steep slope inwards. The surface density scaling $\Sigma_q$ is set to 0.1 g cm$^{-2}$ to match the outer disk CO isotopologue emission, including the upper limit on the C$^{18}$O. The dust surface density profiles of large grains and small grains are parametrized by radially varying the local dust-to-gas ratio and large grain fraction $f_{\rm ls}$. This follows the radii of the SPHERE dust rings as derived above and ALMA millimeter dust ring at 130-160 au, in order to reproduce the millimeter continuum profile and the SED's lack of near-infrared excess (see Supplementary Table \ref{tab:daliparameters}). The resulting surface density profile provides a good match to the SED, the millimeter continuum radial profile, and the general $^{12}$CO and $^{13}$CO emission profiles. The C$^{18}$O emission of the model is consistent with the non-detection. 

In the second step of the modeling process, we parametrize the gaps of the detected protoplanets. The gaps defined by Eqn. \ref{eqn:widegap} for planets 2b and 2c are prescribed at different gap depths $\delta_{\rm gas}$ with Eqn. \ref{eq:fung1} and \ref{eq:fung2} using the known planetary parameters and the $h(r)$ profile derived in step 1, assuming a constant $\alpha$ throughout the disk which regulates the gap depth \citep{SanchezSalcedo2023}.

The inner half-widths are set to 3 and 3.5 $R_H$ and outer half-widths are set to match the observed range in the $\delta\upsilon_{\phi}$ profile in combination with the planet orbital radii and SPHERE dust ring radii (Figure \ref{fig:radialprofiles}): the inner gap by 2c extends from 9 out to the dust ring at 34 au (inner edge is unconstrained by data), the second gap by 2b extends from 40 to the $^{12}$CO peak at 100 au. The gap profile of proposed planet 2d is described by the relations in \citep{Kanagawa2015} for a planet mass of 0.5 $M_{\rm Jup}$ at 130 au based on \citep{Byrne2026}, which is included in the profile. The resulting parameters are listed in Supplementary Table \ref{tab:planets}.

The final surface density profile with the background profile of Eq.~\ref{eqn:brokenlaw} and the three parametrized gaps is shown in  Figure \ref{fig:finaldali}, together with the resulting emission profiles. The depth of the inner gap, created by planet 2c, cannot be constrained at the observed spatial resolution, as indicated by the hatched region in the plot. For the second gap, created by planet 2b, the gap depth is most consistent with a $\delta_{\rm gas}$ value of 5$\cdot10^{-2}$ when comparing the $^{12}$CO profile. With the constraints from Eqn \ref{eq:fung1}, this would correspond to an $\alpha$-viscosity of 4$\times10^{-3}$, although the absolute value of $\alpha$ cannot be fully constrained from the available gap prescriptions in the literature.
The gap full-widths are consistent with 15 and 8.5 $R_H$, respectively, which is a reasonable range for eccentric gaps in simulations. The gap of proposed planet 2d is not sufficiently deep to show up in the optically thick $^{12}$CO emission profile. 

The final surface density corresponds to a low disk gas mass of 1.0 $M_{\rm Jup}$, consistent with the non-detection of the C$^{18}$O emission. On the other hand, when considering the average emitting height $z/r$ of the $^{12}$CO emission derived in the kinematics analysis above and the relation between the $z/r$ and the rotational disk mass \citep{Galloway2025}, the disk mass is suggested to be closer to 0.07$M_*$, or $\sim$100 $M_{\rm Jup}$. This discrepancy could be explained by a process called CO depletion in the outer disk (CO chemical transformation as CO frozen out on the dust grains is transported inwards through radial drift), as suggested in other disks \citep[e.g.][]{Bergin2013,Zhang2021}. The $^{12}$CO brightness temperature indeed drops below 20 K in the outer disk (Supplementary Figure \ref{fig:vel_curve}), so the actual disk mass is possibly higher than constrained by our model.

This modeling procedure demonstrates that the observed gaps in the $^{12}$CO emission can be reproduced by physically driven gap profiles based on planet-disk interaction models of eccentric planets. The general intensity profile suggests that the disk has evolved through MHD-driven disk evolution, where the total width of the gap carved by planet 2b is 8.5 $R_H$ and its depth $\delta_{gas}$=2$\cdot10^{-2}$, pointing to moderate $\alpha$-viscosity considering the relations above. 
Generally, $\alpha$-viscosities in the outer disk constrained from molecular line and dust continuum observations in other disks are found to be between 10$^{-4}$ and 10$^{-3}$ \citep{Rosotti2023}.

\newpage

\bmhead{Acknowledgments}
The authors thank Ewine van Dishoeck and Sijme-Jan Paardekooper for useful discussions regarding the manuscript. NM acknowledges assistance from Allegro, the European ALMA Regional Centre node in the Netherlands. 

\bmhead{Funding}
R.D. acknowledges support from the National SKA Program of China under Grant No. 2025SKA0120100. 
Support for AFI was provided by NASA through the NASA Hubble Fellowship grant No. HST-HF2-51532.001-A awarded by the Space Telescope Science Institute, which is operated by the Association of Universities for Research in Astronomy, Inc., for NASA, under contract NAS5-26555. 

\bmhead{Authors' contributions}
NM reduced and imaged the ALMA data, fit the continuum visibilities, and performed the thermal-chemical modeling. NM and JS wrote the main text and made the main figures.
JS and AFI analyzed the disk gas kinematics of the ALMA data. 
RD and BB contributed to the discussions on planet-disk interaction models and implications for giant planet formation.
JB did the new fitting of the rings in the SPHERE image.
RC extracted the astrometry of the planets.
CL did the new orbital fitting of the planet orbits and RC provided valuable input.
CG and MK contributed to the discussions on the interpretation of the SPHERE data and the previous protoplanet detections.
All authors contributed with feedback on the text in the manuscript.

\bmhead{Competing interests} The authors declare no competing interests.

\bmhead{Availability of data and materials}
This paper makes use of the following ALMA data: ADS/JAO.ALMA\#2025.1.00433.S. ALMA is a partnership of ESO (representing its member states), NSF (USA) and NINS (Japan), together with NRC (Canada), NSTC and ASIAA (Taiwan), and KASI (Republic of Korea), in cooperation with the Republic of Chile. The Joint ALMA Observatory is operated by ESO, AUI/NRAO and NAOJ. The data sets generated and analysed during the current study will be made available in a Zenodo repository upon acceptance.

\bmhead{Code availability} 
All the code for the analysis and the generation of all the figures are available upon request. The raw ALMA data and pipeline reduction scripts for dataset 2025.1.00433.S are publicly available through the ALMA Science Archive (\url{https://almascience.eso.org/aq/}). The ALMA data reduction software CASA is publicly available (\url{https://casa.nrao.edu/}).
The \texttt{discminer} channel-map modeling and analysis framework \citep{Izquierdo2021} is available through GitHub (\url{https://github.com/andizq/discminer}). The DALI code is private but available upon request from N. van der Marel. 

\bmhead{Supplementary information}
Supplementary Information is available for this paper.

\bmhead{Correspondence and requests for materials}
should be addressed to Nienke van der Marel.

\bmhead{Reprints and permissions information} is available at http://www.nature.com/reprints.
\newpage

\pagenumbering{gobble}
\bibliography{sn-bibliography}

\newpage

\section*{Supplementary information}
\begin{figure}[!ht]
    \centering
\includegraphics[width=0.4\textwidth]{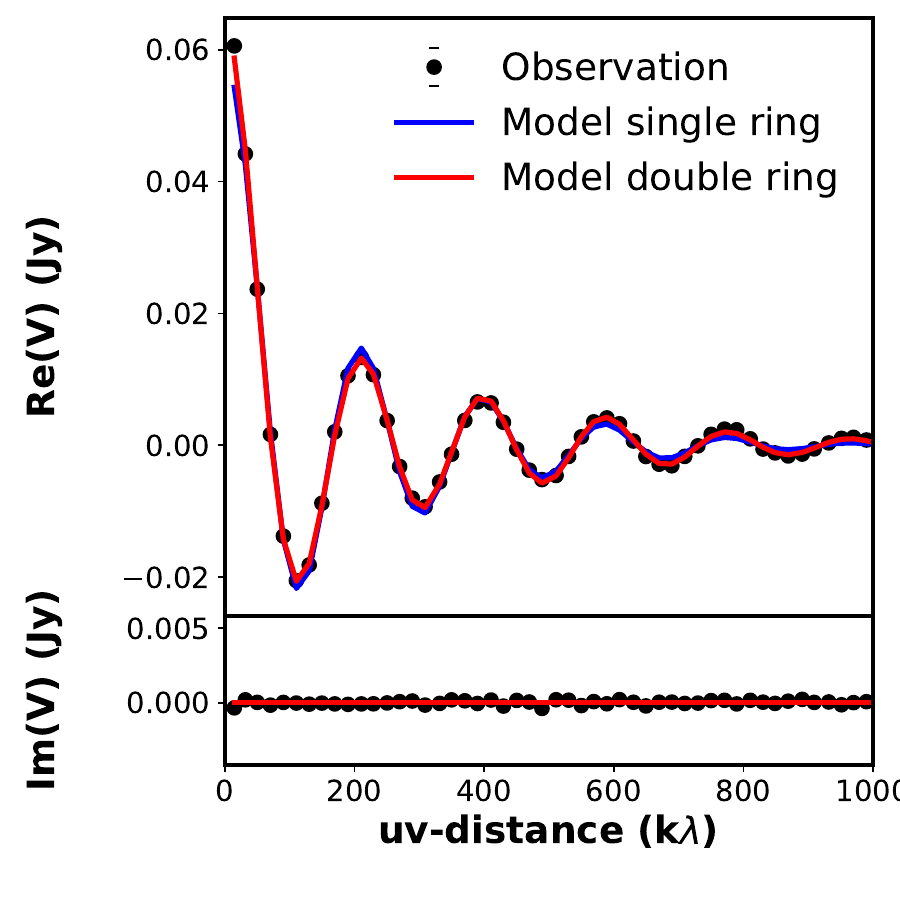}
\includegraphics[width=0.55\textwidth]{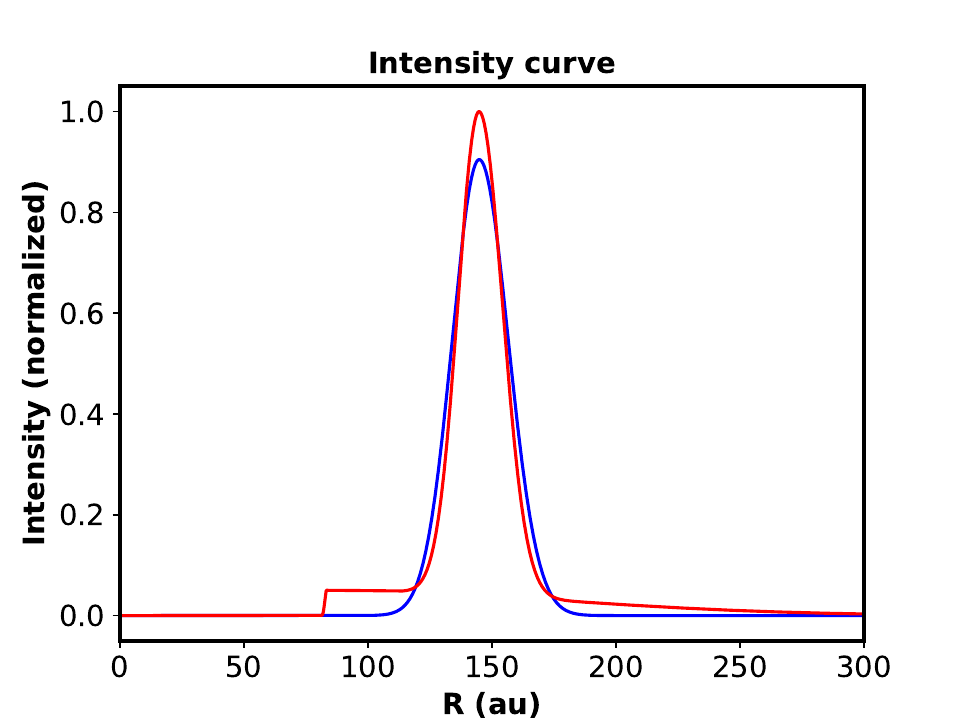}
    \caption{Best-fit model of the continuum visibility fitting. The left panel shows the binned and deprojected visibilities of the observation, compared with the best-fit model of the single ring (in blue) and the double ring (in red). The normalized radial profile of the models is shown in the right panel, normalizing both profiles to the double ring profile. Whereas the fitting of both models converges, the double ring model fully reproduces the observed flux at the shortest baselines, but the single ring model does not.}
    \label{fig:vismodel}
\end{figure}

\begin{figure}[!ht]
    \centering  
    \includegraphics[width=\textwidth]{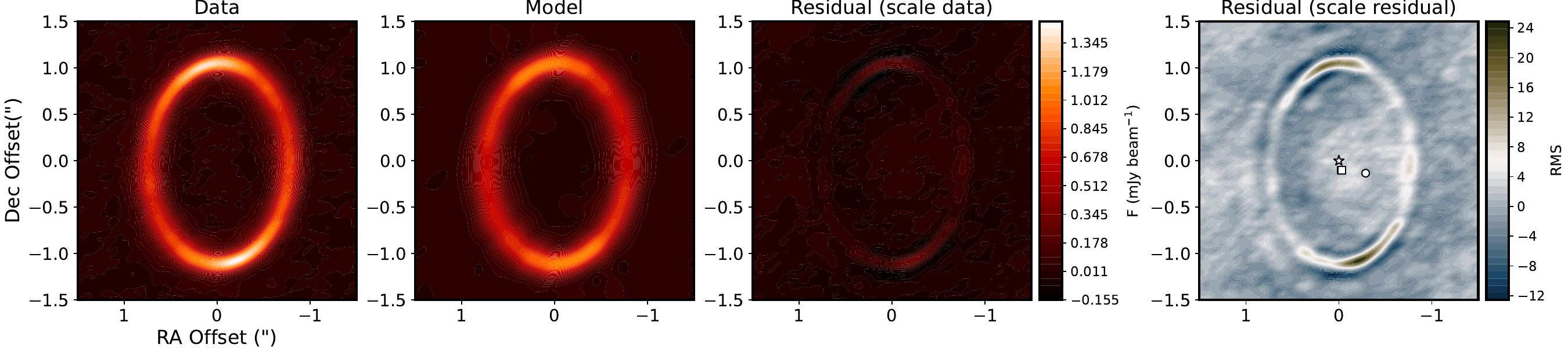}   
    \caption{Images of the best fit from the visibility fitting of the double ring model, mapped onto the observed visibilities and imaged with the same parameters as the data. a) Original data. b) Best-fit model at the same color scale as the data. c) Residual at the same color scale as the data. d) Residual divided by rms ($\sigma$=0.018 mJy beam$^{-1}$) with the colors rescaled to its own minimum and maximum. The locations of the planets and star are indicated.}
\label{fig:imagevismodels}
\end{figure}

\begin{figure}[!ht]
    \centering
    \includegraphics[width=\textwidth]{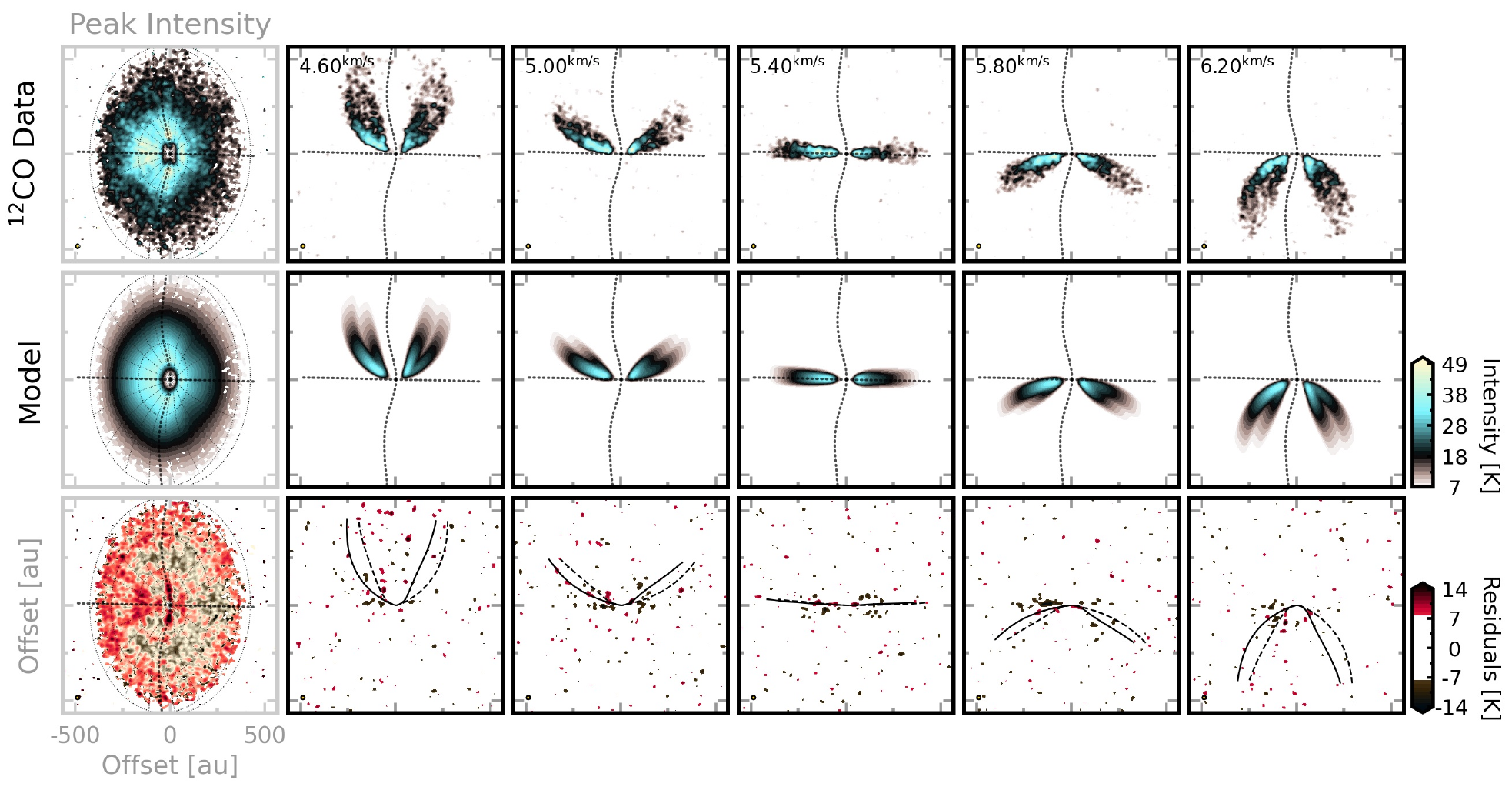}
    \caption{$^{12}$CO channel maps gallery. Top: Observations. Middle: \texttt{discminer} model channel maps. The dotted lines trace the disk's minor and major axes. Bottom: Residuals between data and model channel maps with values masked below $3\,\sigma$ ($\sigma=2.6\,$K). The solid and dashed lines indicate the isovelocity contours extracted from the upper and lower surface velocities at the value of the corresponding velocity channel. The leftmost columns show the Gaussian peak brightness temperature maps calculated using the Rayleigh-Jeans approximation for the data, the model, and the residuals.}
    \label{fig:dm_channels}
\end{figure}

\begin{figure}[t]
    \centering
    \includegraphics[width=\linewidth]{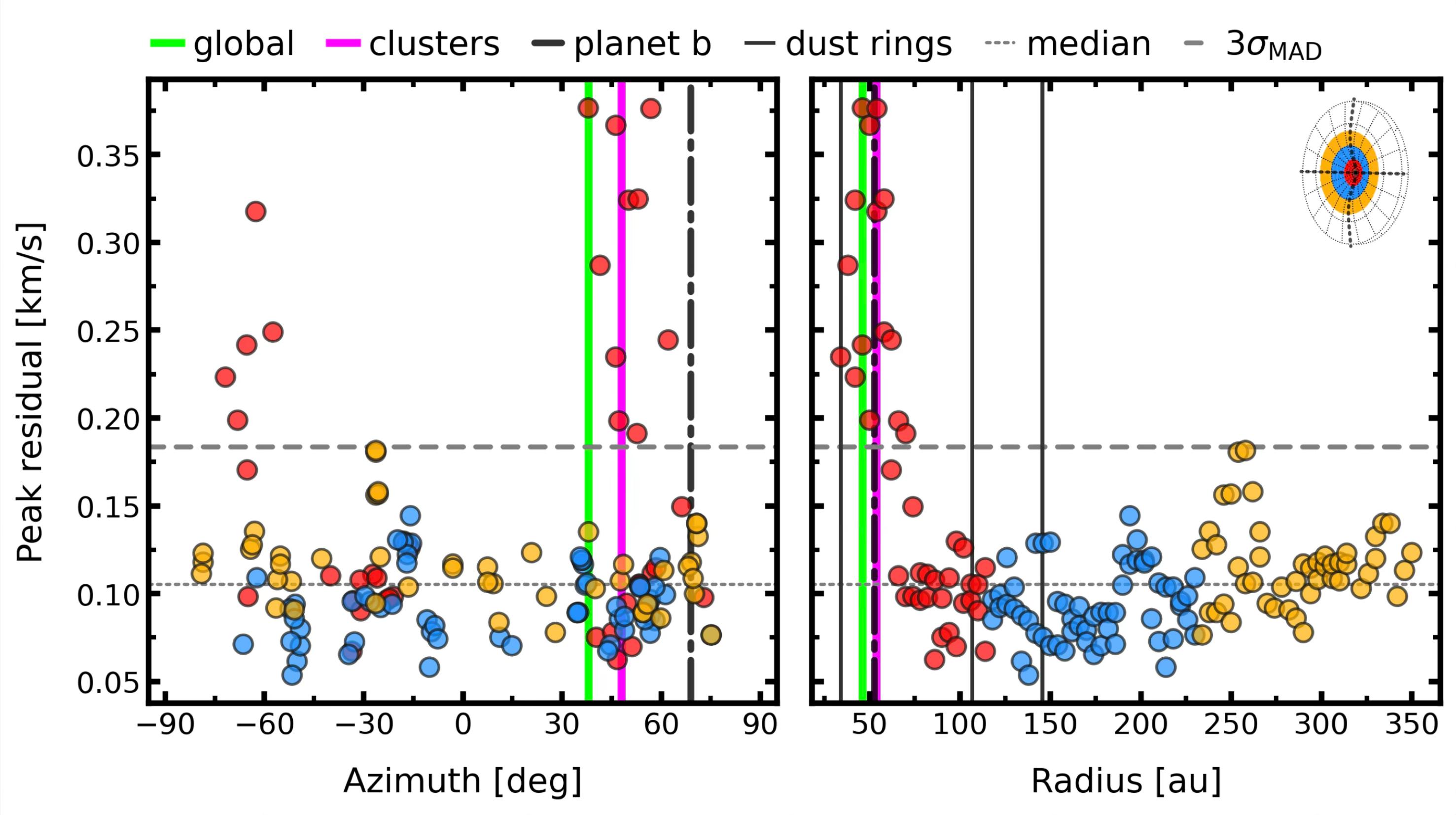}
    \caption{Distribution of peak centroid residual points in azimuth and radius. The points are color-coded to reflect their radial positions within the disk shown in the top-right corner. The locations of the global peak (lime), weighted-mean cluster centers (magenta), \wisb (black, deprojected from current on-sky position), and continuum ring (grey) are highlighted by colored vertical lines. Peak residual points $p$ are considered significant once they surpass a $3\sigma_{\rm MAD}$ threshold in terms of median absolute deviation (${\sigma_{\rm MAD}=1.4826\times \mathrm{median}(|p-\mathrm{median}(p)|)}$), as shown by the horizontal dashed line \citep[see][for details]{Izquierdo_ea_2026b}. Note that whether a cluster of peak residual points is deemed significant is assessed in terms of the cluster's velocity variance via the K-means cluster algorithm, see Figure\,\ref{fig:phi_clusters}.}
    \label{fig:folded_map_peaks}
\end{figure}

\begin{figure}[t]
    \centering
    \includegraphics[width=\linewidth]{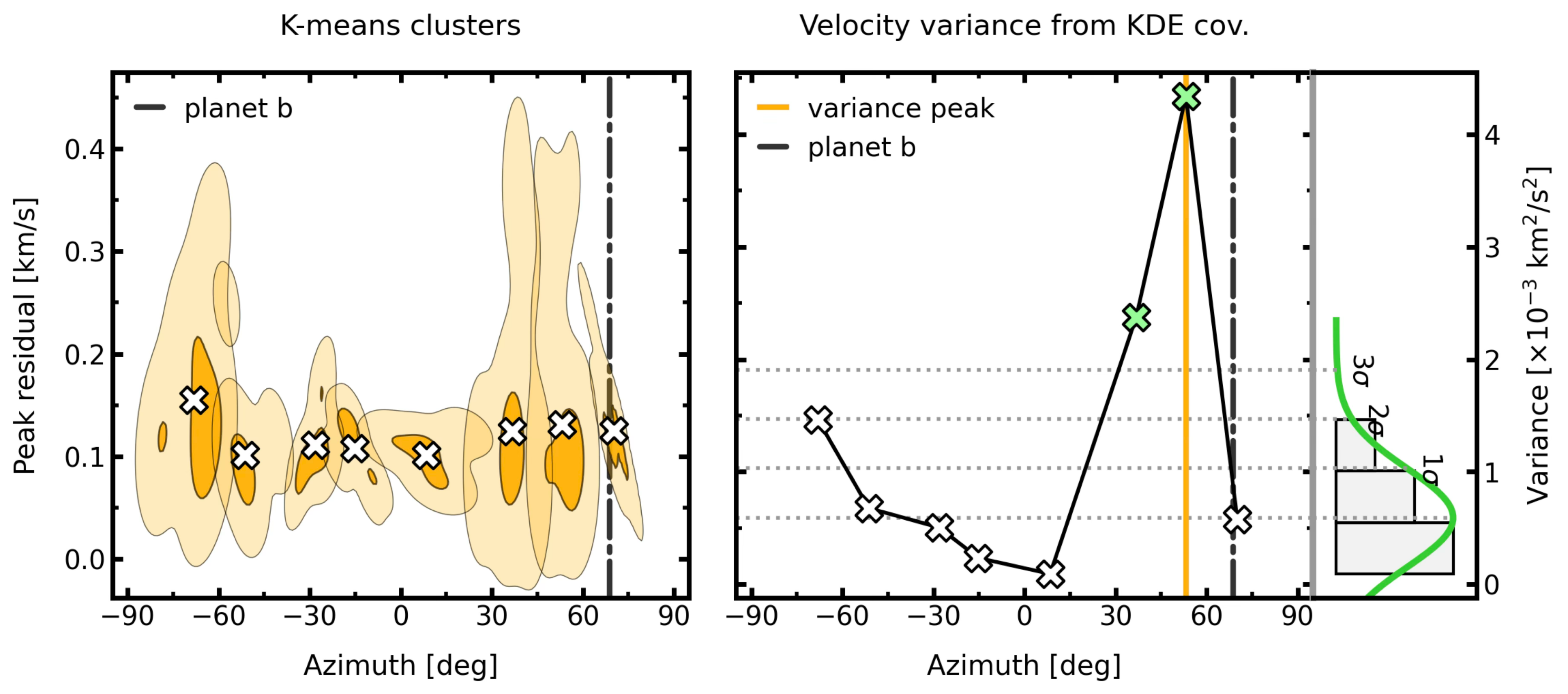}
    \caption{Azimuthal K-means cluster statistics. Left: The yellow contours enclose 33\% and 95\% of the peak residuals for each azimuthal cluster. The white crosses are the cluster centers determined by the K-means algorithm. Right: The spectral variance of peak residuals is shown for each cluster. In the panel attached on the right, we show 1, 2, and 3$\sigma$ significance thresholds. Clusters above $3\sigma$ are considered significant and are highlighted by green crosses at $37\degree$ and $53\degree$ with $3.7\sigma$ and $7.8\sigma$ significance, respectively. These can be attributed to the planet b's perturbation, its azimuthal location indicated by the vertical black line. The yellow line highlights the azimuthal location of the most significant cluster. Note: While the cluster at $\sim-70\deg$ contains a few peak residual points above $3\sigma_{\rm MAD}$ (see left panel of Ext. Fig.~\ref{fig:folded_map_peaks}), its velocity variance (spread and number of points) is not significant ($\sim2\sigma$).}
    \label{fig:phi_clusters}
\end{figure}

 \begin{figure}[!ht]
    \centering
    \includegraphics[width=0.9\linewidth]{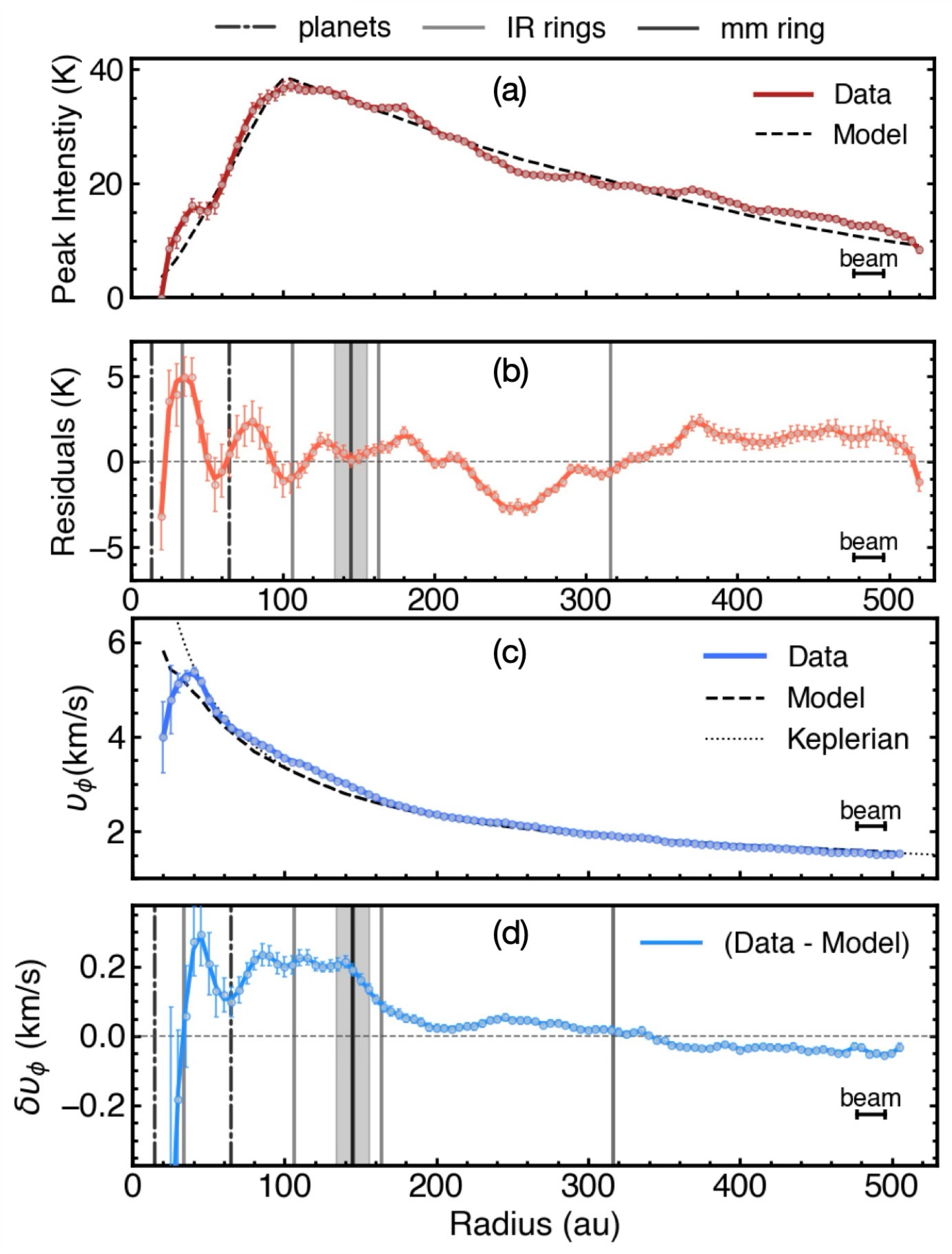}
    \caption{$^{12}$CO radial profiles: (a) Peak brightness temperature for data and model. (b) Temperature residuals. (c) Rotational velocity of data, model,  and background Keplerian. (d) Deviations from Keplerian rotation.  The vertical gray and black solid lines show the peaks of the $H$-band IR rings and dust continuum ring, respectively. The black dashed-dotted lines show the radial positions of the planets \wisa b\,\&\,c. The continuum ring aligns with a negative $\delta\upsilon_\phi$-gradient, indicative of a gas-pressure bump.}
    \label{fig:vel_curve}
\end{figure}

\begin{figure}[!ht]
    \centering
    \includegraphics[width=0.9\textwidth]{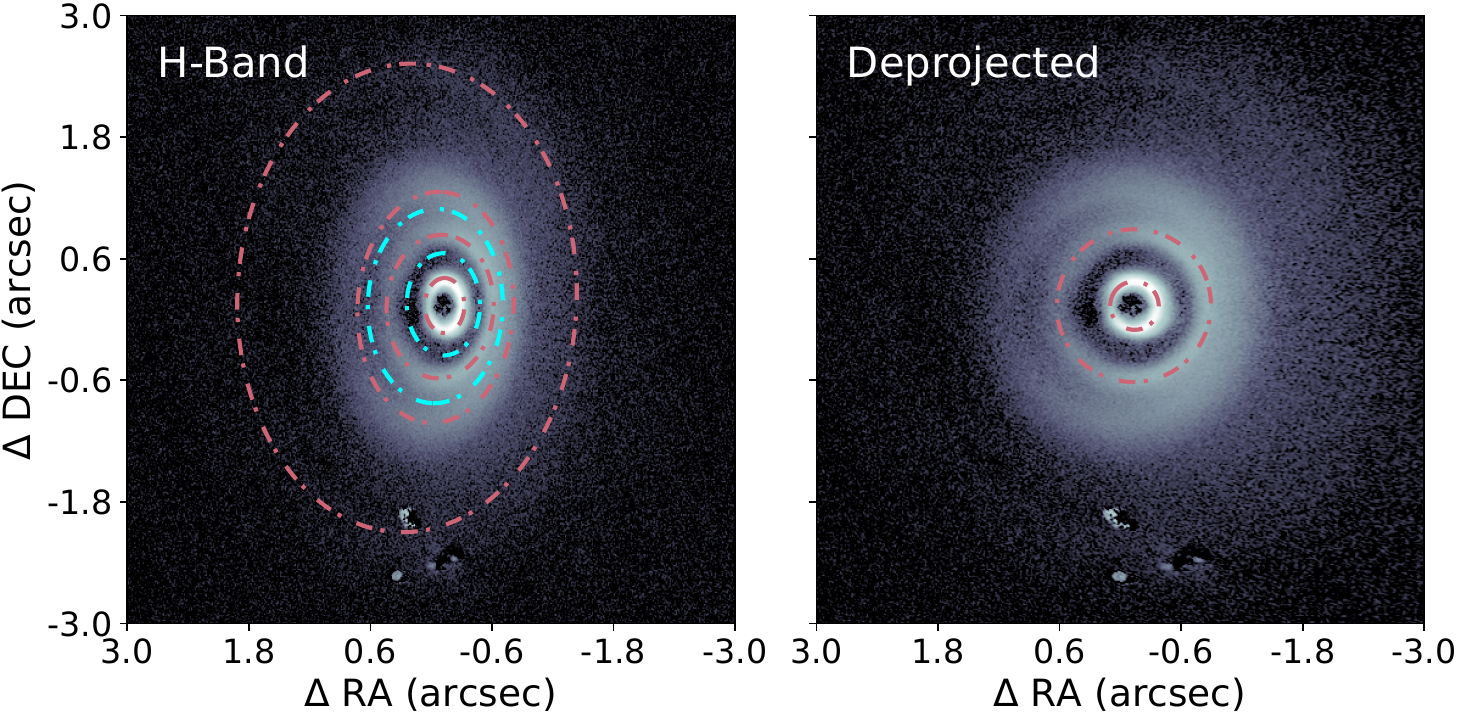}   
    \caption{$H$-band $Q_\phi$ image of \wisa with fitted ellipses overlaid (left) and the corresponding deprojected image (right). Rings are shown in rose and gaps in cyan. Both panels use the same logarithmic stretch.}
    \label{fig:ellipsefits}
\end{figure}

\begin{figure}[!ht]
    \centering
    \includegraphics[width=0.45\textwidth]{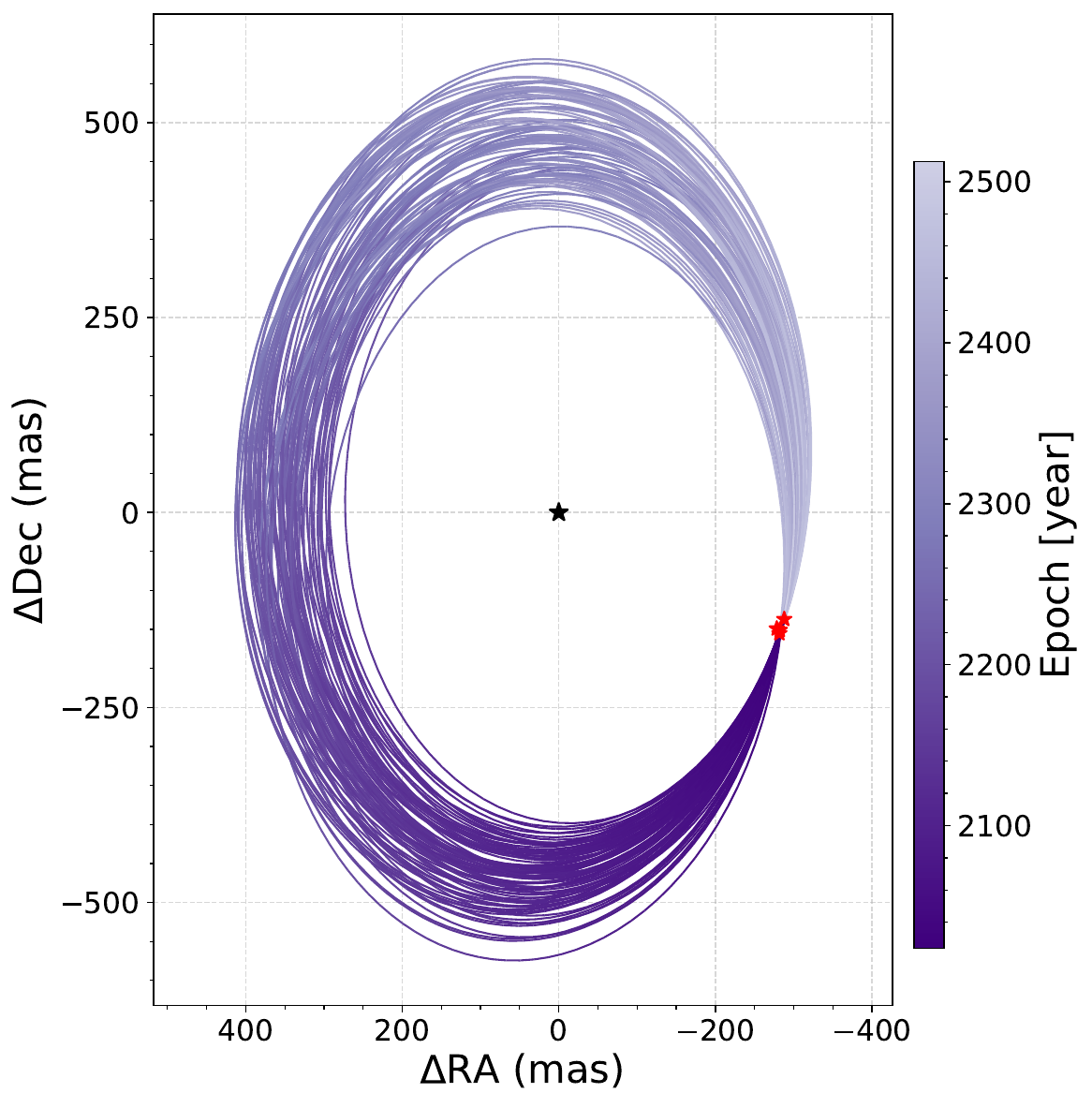} 
        \includegraphics[width=0.45\textwidth]{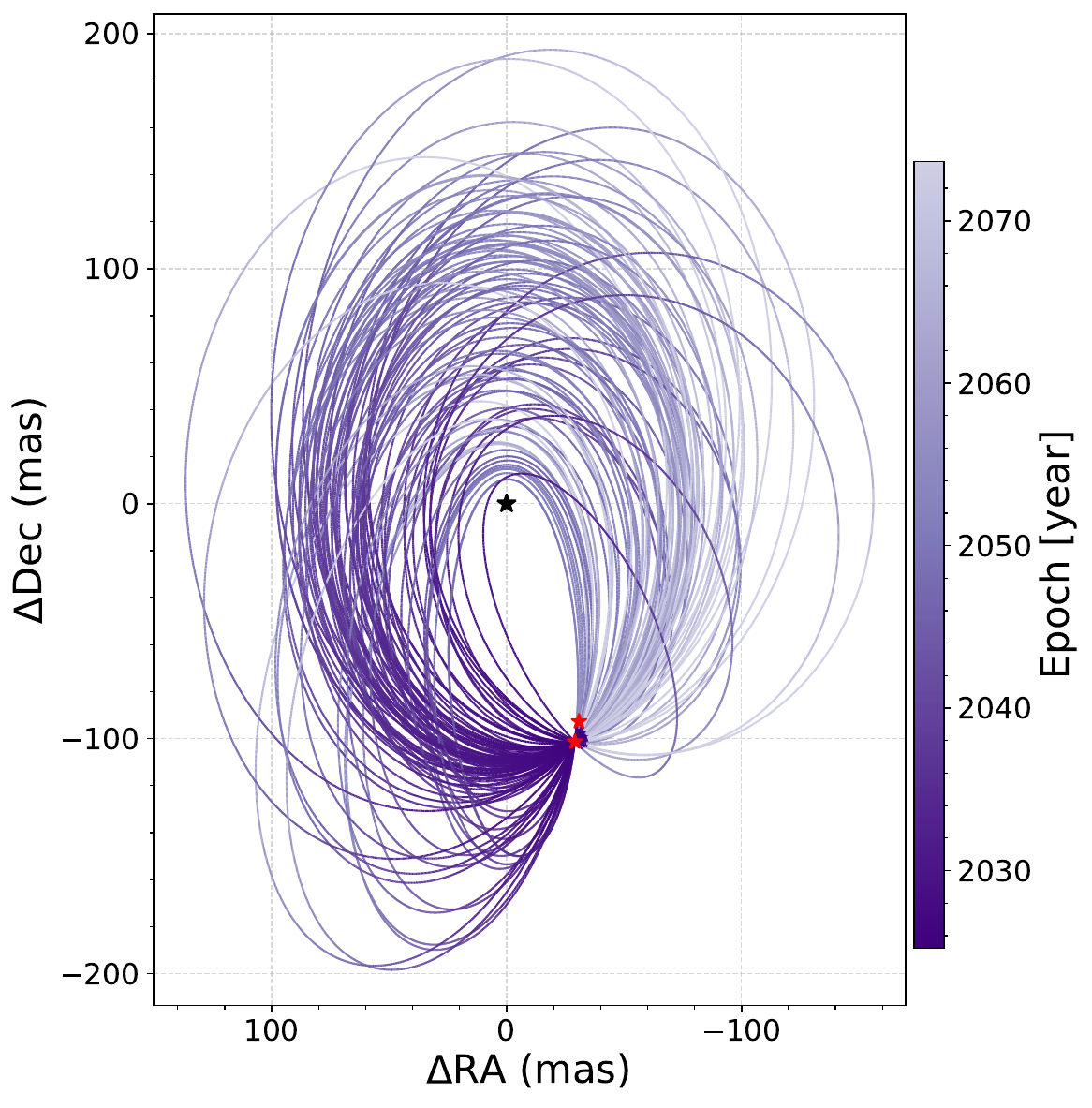} 
    \caption{Orbital fits of protoplanet astrometry. Left: Updated orbital families of \wisb in relation to the central host marked by the black star. Right: Orbital families of \wisc in relation to the central host marked by the black star. Astrometric points are denoted by the red stars along the projected orbital motion curves.} 
    \label{fig:WISPIT_2b_orbit}
\end{figure}

\clearpage
\newpage

\begin{table*}[!ht]
    \centering
    \caption{ALMA observations of \wisa}
    \label{tbl:obs}    
    \begin{tabular}{lllllll}
    \hline
         Date & Antennas & Baseline & Time on & PWV & Bandpass/flux  & Gain calibrator  \\
         &&range&source&&calibrator&\\
         &&(m)&(min.)&(mm)&&\\
         \hline
         2025-10-23 & 50 & 34-8283 & 45 & 0.6 & J1924-2914 & J1920-0236 \\
         \hline
         2025-12-18 & 47 & 15-952 & 11 & 2.7 & J1924-2914 & J1954-1123 \\
         2025-12-21 & 45 & 15-784 & 11 & 2.2 & J1924-2914 & J1954-1123 \\
         \hline         
    \end{tabular}
\end{table*}

\begin{table}[!ht]
    \centering
    \caption{Best fit parameters visibility modeling}
    \label{tbl:galariofit}    
    \begin{tabular}{ll|cc}
    \hline
    && Single ring & Double ring \\
         Param. & Unit &  model & model \\
         \hline
$\log I_1$ & (arb.) & 9.383$\pm$0.001 & 9.408$\pm$0.001\\
$r_{c1}$ & (") & 1.090 $\pm$ 0.001 & 1.090$\pm$0.001\\
$r_{w1a}$ & (") & 0.0818 $\pm$ 0.0005 & 0.065$\pm$0.001\\
$r_{w1b}$ & (") & 0.0879 $\pm$ 0.0005 & 0.071$\pm$0.001\\
$\log I_2$ & (arb.) & & 8.13$\pm$0.01 \\
$r_{c2}$ & (") &  & 0.63$\pm$0.02 \\
$r_{w2a}$ & (") &  & 0.005 $\pm$ 0.004$^a$\\
$r_{w2b}$ & (") &  & 0.70 $\pm$ 0.02\\
\hline
$i$ & ($^{\circ}$) & 45.65 $\pm$ 0.01 & 45.53 $\pm$0.03\\
PA & ($^{\circ}$) & 178.53 $\pm$ 0.05 & 178.52 $\pm$0.03\\
dRA & (") & 0.0020 $\pm$ 0.0002 & 0.0017$\pm$0.0002\\
dDec & (") & 0.0080 $\pm$ 0.0002 & 0.0077$\pm$0.0002\\
         \hline
    \end{tabular}
    $^a$ The parameter $r_{w2a}$ converges towards an upper limit.
\end{table}

\setlength{\tabcolsep}{6pt} 

\begin{table}
\centering
{\renewcommand{\arraystretch}{1.5}
 \caption{List of best-fit parameters for our discminer channel-map model of the $^{12}$CO emission line from the disk of Wispit~2.}
  \label{table:attributes_parameters}
\begin{tabular}{ l|l } 

\toprule
\toprule
Attribute &  \multicolumn{1}{c}{Best-fit discminer parameters} \\
\hline

Orientation & $i=46.0^\circ$ \quad $\rm{PA}^1 = 178.7^\circ$ \quad $x_c=27.6$\,mas \quad $y_c=-0.3$\,mas \\

\hline

Velocity & $M_\star=1.37$\,M$_\odot$ \quad $\upsilon_{\rm LSRK}=5.44$\,km\,s$^{-1}$  \\

\hline

Upper surface$^{2}$ & $z_0 = 49.6$\,au \quad $p=2.61$ \quad $R_t = 120.1$\,au \quad $q=0.98$ \\

Lower surface & $z_0 = 21.6$\,au \quad $p=2.77$ \quad $R_t = 155.3$\,au \quad $q=1.11$ \\

\hline
Peak intensity$^{3}$ & $I_0 = $98.2\,mJy\,pix$^{-1}$ \quad $p_1=0.09$ \quad $p_2=-1.28$  \\
               & $q=0.64$\quad $R_{\rm break}=99.3$\,au \quad $R_{\rm out}=565.3$\,au \\

Line width & $L_{w0} = 0.57$\,km\,s$^{-1}$ \quad $p=-0.84$ \quad $q=0.05$  \\ 

Line slope &  $L_{s0} = 2.16$ \quad $p=0.17$  \\

\bottomrule

\end{tabular}

\vspace{0.5cm}

\justifying
{\noindent \textbf{Note.} The best-fit model parameters are the mean of 16th and 84th percentiles of the posterior distribution of the final 10\% of the walkers. The composite model peak intensity is set to zero for radii beyond $R_{\rm out}$. For a detailed parameterization of the model see \cite{izquierdo_ea_2025}. The MCMC uncertainties on the reported parameters are on the order of 0.1\,\%, but can increase by a factor of $\sim10$ due to spatially correlated noise, as shown by \cite{Hilder_ea_2025}.\\
$^{1}$ The PA is measured counter-clockwise from the North to the disk's red-shifted major axis. \\
$^{2}$ Both surfaces are modeled as exponentially tapered power laws.  \\
$^{3}$ A double power law was prescribed for the model intensity profile to account for the observed gas cavity.}
  
  }
\end{table}

\begin{table*}[!h]
\caption{Geometric parameters derived from ellipse fitting of the deprojected $H$-band scattered light image.}
\label{tab:constrained_summary}
\centering
\def\arraystretch{1.2}
\setlength{\tabcolsep}{8pt}
\begin{tabular*}{\textwidth}{@{\extracolsep{\fill}}lll}
\hline\hline
Structure & Separation (au) & Eccentricity \\
\hline
IR Ring 2 & $106.647 \pm 0.122$ & $0.132 \pm 0.013$ \\
IR Ring 3 & $33.805 \pm 0.206$ & $0.128 \pm 0.070$ \\
\hline
\end{tabular*}
\end{table*}

\begin{table}[!ht]
    \centering
    \caption{Parameter values of the density structure in the final DALI model}
    \label{tab:daliparameters}    
    \begin{tabular}{c|l|c}
    \hline
         Radial structure & $r_{\rm sub}$ (au) & 0.06 \\
         & $r_{\rm out}$ (au) & 600 \\ 
         & $r_{\rm ring3}$ (au) & 34 \\
         & $r_{\rm ring2}$ (au) & 107 \\
         & $r_{\rm mmdust}$ (au) & 130-160 \\
         \hline
         Gas surface density& $r_t$ (au) & 150 \\
         & $\Sigma_t$ (g cm$^{-2}$) & 0.1 \\
         & $t$ & 1.5 \\
         \hline
         Vertical structure & $h_c$ (rad) & 0.12 \\
         & $r_c$ (au) & 150 \\
         & $\psi$ & 0.5 \\
         & $f_{\rm ls}$ ($< r_{\rm mmdust}$) & 0.0 \\
         & $f_{\rm ls}$ ($> r_{\rm mmdust}$) & 0.85 \\
         & $\chi$ & 0.2 \\
         \hline
         Dust-to-gas ratio & $\delta_{\rm dust}$ ($<r_{\rm ring3}$) & 10$^{-20}$ \\
         & $\delta_{\rm dust}$ ($<r_{\rm mmdust}$) & 10$^{-1}$ \\
         & $\delta_{\rm dust}$ ($r_{\rm mmdust}$ range) & 0.5 \\
         & $\delta_{\rm dust}$ ($>r_{\rm mmdust}$ range) & 10$^{-1}$ \\
         \hline
   \end{tabular}
\end{table}

\begin{table}[!ht]
\centering
\caption{Gap parameters of the final DALI model}
\label{tab:planets}
\begin{tabular}{llllllll}
\hline
    Planet & $r_p$ & $m_p$ & $q$ & $h$ & $w_{\rm in,half}$ & $w_{\rm out,half}$ &$r_H^a$ \\
    & (au) & ($M_{\rm Jup}$) & & & (au) & (au) & (au) \\ 
    \hline
    2c & 15 & 9.0 & 6.4$\times10^{-3}$ & 0.037 & 5.4 (3$R_H$) & 22 (12$R_H$) & 1.9 \\
    2b & 65 & 5.0 & 3.6$\times10^{-3}$ & 0.074 & 25 (3.5$R_H$) & 35 (5$R_H$) & 6.9 \\
    2d$^b$ & 130 & 0.5 & 3.5$\times10^{-4}$ & 0.10 & 13 (2$R_H$) & 13 (2$R_H$) & 6.4 \\
\hline    
\end{tabular}
$^a$ Hill radius $r_H$ is defined as $r_p\times(q/3)^{1/3}$, see text.\\
$^b$ The existence of planet 2d is still speculative at time of writing, and is just included to demonstrate that the observed CO profile is consistent with its potential presence.
\end{table}

\end{document}